\documentclass[twocolumn]{aastex631}
\makeatletter
\let\frontmatter@title@above=\relax
\makeatother

\newcommand{\psr}{PSR\,B1508+55}
\usepackage{amsmath}
\usepackage{graphicx}

\begin{document}

\title{Identifying pulsar candidates in interferometric radio images using scintillation}

\author{Jitendra Salal}
\affiliation{National Centre for Radio Astrophysics, Post Bag 3, Ganeshkhind, Pune, 411007, India} 

\author{Shriharsh P. Tendulkar}
\affiliation{Department of Astronomy and Astrophysics, Tata Institute of Fundamental Research, Mumbai, 400005, India}
\affiliation{National Centre for Radio Astrophysics, Post Bag 3, Ganeshkhind, Pune, 411007, India}
\affiliation{CIFAR Azrieli Global Scholars program, CIFAR, Toronto, Canada}

\author{Visweshwar Ram Marthi}
\affiliation{National Centre for Radio Astrophysics, Post Bag 3, Ganeshkhind, Pune, 411007, India}

\begin{abstract}

Pulsars have been primarily detected by their narrow pulses or periodicity in time domain data. Interferometric surveys for pulsars are challenging due to the trade-off between beam sensitivity and beam size and the corresponding tradeoff between survey sensitivity (depth), sky coverage, and computational efforts. The detection sensitivity of time-domain searches for pulsars is affected by dispersion smearing, scattering, and rapid orbital motion of pulsars in binaries. We have developed a new technique to select pulsar candidates in interferometric radio images by identifying scintillating sources and measuring their scintillation bandwidth and timescale. Identifying likely candidates allows sensitive, focused time-domain searches, saving computational effort.  Pulsar scintillation is independent of its timing properties and hence offers a different selection of pulsars compared to time-domain searches. Candidates identified from this method could allow us to find hard-to-detect pulsars, such as sub-millisecond pulsars and pulsars in very compact, highly-accelerated binary orbits. We use uGMRT observations in the fields of PSR\,B1508+55, PSR\,J0437$-$4715, and PSR\,B0031$-$07 as test cases for our technique. We demonstrate that the technique correctly differentiates between the pulsar and other non-scintillating point sources and show that the extracted dynamic spectrum of the pulsar is equivalent to that extracted from the uGMRT phased array beam. We show the results from our analysis of known pulsar fields and discuss challenges in dealing with interference and instrumental effects.\\

\end{abstract}

\section{Introduction} \label{sec:intro}

Pulsars, highly magnetized rotating neutron stars resulting from the remnants of massive stars, were first discovered in 1967 as a series of periodic pulses \citep{hewish}. These pulses exhibit varying periods, ranging from milliseconds to tens of seconds, with most pulsars falling within the range of 0.1 to 10 seconds \citep{manchester}. The distribution of pulsar periods is biased due to the predominant use of the Fast Fourier Transform (FFT) for their detection \citep{cameron}.\\

The FFT method, however, is less sensitive to slow-rotating pulsars due to the presence of red noise, thereby resulting in a higher detection rate for faster-rotating pulsars. To address this bias, recent advancements in computing power have facilitated the introduction of the Fast Folding Algorithm \citep[FFA; ][]{cameron}. The FFA allows for the detection of slower pulsars, thereby mitigating the bias in the period distribution of the overall pulsar population. By implementing the FFA and other improved computational techniques, researchers have made significant progress in uncovering a larger number of pulsars with diverse rotation periods \citep{handbook}. These advancements have contributed to a more comprehensive understanding of pulsar populations and their characteristics, further enriching the field of pulsar research.\\

Still, algorithms like FFT or FFA are not inherently sensitive to all types of pulsars, particularly those found in binary systems, due to the significantly large parameter space involved. For isolated pulsars, the search space includes dispersion measure, period ($P$), period derivative ($\dot{P}$), second-period derivative ($\ddot{P}$). For pulsars in binary systems, an additional searches in acceleration and jerk space is necessary, further increasing the computational resources required. However, it is important to note that even these searches may not be sufficient to identify pulsars in very compact binary systems, as additional motion derivatives come into play.\\

Although more than 3,000 pulsars have been discovered \citep{manchester}, population synthesis simulations suggest that there $\sim$ 120000 isolated pulsars present in our galaxy \citep{faucher}. If we include binary pulsars, this number would be even larger. Comparing the number of detected pulsars to the total number estimated to exist in the Galaxy, it becomes apparent that less than 3$\%$ of them have been identified so far. This low detection fraction can be due to multiple reasons: propagation effects such as scattering and complex binary motions, as well as low pulse modulation due to a chance alignment of the rotational and magnetic axis \citep{young}.\\

One approach to overcome some of the limitations associated with periodic searches is to detect putative pulsar candidates in radio images. With this selection, we can focus our computational efforts more efficiently. Additionally, radio imaging is not affected by dispersion measure smearing, scattering, or the orbital motion of spin periods, making it a robust technique for detecting elusive pulsars such as sub-millisecond pulsars, pulsar-black hole systems, and pulsars located near the Galactic center.\\

\subsection{Previous Work for Identifying Pulsar Candidates}

Steep spectra, compact sizes, and circular polarization have been used to identify pulsar candidates and focus observational efforts. The first-ever millisecond pulsar, J1939+2134, was discovered in radio continuum images as an unusual compact source with a steep spectrum \citep{backer}. However, not all searches for pulsations from steep spectrum sources may be successful for various reasons  \citep{Kaplan_2000} including contamination of the sample by extragalactic interlopers. The follow-up of Fermi gamma-ray sources to identify millisecond pulsars using GMRT and Parkes radio search results in the discoveries of millisecond pulsars \citep{frail2018, Bhattacharyya,camilo}. More recently, the Australian Square Kilometer Array Pathfinder telescope (ASKAP) has successfully detected another pulsar in radio images by leveraging its high polarization characteristics \citep{kaplan2019}. Furthermore, the identification of a circularly polarized variable radio source in the Large Magellanic Cloud (LMC) suggests the potential for detecting new pulsars in this region \citep{Wang_2022}.\\

A recent study by \citep{dai} explores the identification of potential pulsar candidates based on their scintillation time scales and bandwidths, which have to be comparable to the time and frequency resolution of the instrument. The authors employed a variance imaging technique, which theoretically has the potential to detect pulsar candidates. However, in practice, they found that the sensitivity of this method is limited due to background noise. The variance-based approach relies on the principle that pulsars exhibit scintillation, leading to high variance in flux measurements, while other sources that do not scintillate show minimal variance. The variance measures the noise properties across observation time and bandwidth, but does not utilize the structured patterns and correlations created by scintillation.\\

In this work, we build on this idea to leverage the scintillation structure derived from generic visibility data to identify pulsar candidates. By incorporating the detailed scintillation information, including its frequency and time characteristics, we aim to enhance the sensitivity and accuracy of the pulsar identification process.\\

The organization of this paper is as follows: In Subsection \ref{sec:scintillation}, we briefly discuss scintillation due to the interstellar medium. In Section~\ref{sec:gmrt_data} and Section~\ref{sec:data_analysis}, we explain the observations, data processing and analysis techniques. In Sections \ref{sec:results} and \ref{sec:discussion}, we discuss the results, and discuss applications to survey data, challenges, and limitations of this method. We conclude the paper in Section \ref{sec:conclusion} with a look towards future applications.\\

\subsection{Scintillation}
\label{sec:scintillation}

When a plane wavefront passes through turbulent interstellar plasma, its phase (and sometimes amplitude) becomes distorted. As this distorted wavefront reaches the Earth, it interferes with itself, creating an interference pattern. When the Earth/observer traverses this interference pattern, it leads to flux variations over time and frequency due to the relative motion between the pulsar, interstellar medium (ISM), and Earth. The observed regions of constructive and destructive interference in the time and frequency plane (known as scintles) are randomly distributed. The timescale and frequency scale at which these scintles decorrelates are known as the scintillation timescale and bandwidth, respectively.\\

Pulsars scintillate as their angular size ($\sim$ 10\,km/1kpc $\sim$ $10^{-15}$\, radian) is smaller than the angular size scale of interstellar turbulence ($\sim$ $10^{4}$\,km/1kpc $\sim$ $10^{-12}$\, radian). This distinguishes them from the other point sources, such as quasars which have much larger  angular size \citep[$\sim \mathrm{1\,pc/1\,Gpc}\sim10^{-9}\,\mathrm{radian}$; ][]{narayan} .\\

The scintles are distributed randomly throughout the dynamic spectrum and their sizes change gradually with frequency. The scintillation bandwidth refers to the half-width at half-maximum of the auto-correlation in the frequency domain, while the scintillation timescale is the half-width at 1/$e$ of the auto-correlation along the time axis. The average size of a scintle can be determined by performing auto-correlation on the dynamic spectrum. In the case of pulsars, typical diffractive scintillation timescales and bandwidths are on the order of minutes and MHz, respectively, at frequencies $\sim$1\,GHz  \citep{doi:10.1146/annurev.aa.28.090190.003021}.\\

When considering a pulsar located at a distance $D$ from the Earth, we can estimate the scintillation timescale and bandwidth of diffractive interstellar scintillation (DISS) observed at a reference frequency $\nu$. This estimation assumes the presence of a thin scattering disc at $D/2$, characterized by electron density fluctuations following a Kolmogorov spectrum and an effective velocity $V_{eff}$.
\begin{equation}
    \begin{split}
        &\tau_{s}\propto \nu^{6/5}D^{-3/5}V_{\mathrm{eff}}^{-1},\\ &\nu_{s}\propto \nu^{22/5}D^{-11/5}
    \end{split}
\end{equation}

It is worth noting that the DISS bandwidth changes faster with frequency and distance compared to the DISS timescale. Since scintillation timescale and bandwidth decrease with increasing distance,  it is crucial to allocate sufficient bandwidth and integration time to achieve the required sensitivity for detecting scintillation in far away pulsars.\\

Scintillation studies are most often performed using dedicated phased array beams from interferometric telescopes or single beams from single-dish telescopes. Here we demonstrate a technique to identify scintillation in all the point sources of an interferometric image. From a calibrated image, we identify all the point sources and construct dynamic spectra for all the point sources. The dynamic spectra are then used to measure scintillation parameters. We use observations from three known pulsars to demonstrate our technique and verify it with the dynamic spectrum measured from the standard phased array beam.\\

\section{Observations} 
\label{sec:gmrt_data}
The upgraded Giant Metrewave Radio Telescope \citep[uGMRT; ][]{yashwant} is composed of 30 antennae with a 45\,m aperture. These antennae are distributed across a 25\,km region, resulting in higher-resolution radio images. Within this configuration, 14 antennae are concentrated at the center, covering a compact region of 1.1\,km. The remaining antennae are arranged in a Y-shaped array, maximizing the coverage of $uv$ (baseline) measurements as the Earth rotates.\\

We used archival observations of \psr\ \citep[project code DDTB273; ][]{marthi} due to its brightness, with a signal-to-noise ratio (SNR) of approximately 533 at a frequency of 650\,MHz. The observations consist of 3 scans, and we selected the first scan with a duration of 46 minutes for further analysis. The target scans are interleaved with phase calibrator scans of J1400+621 with three scans. No flux calibrator was observed during these scans. We used a phase calibrator as a flux calibrator and set its flux density at 6.74\,Jy based on the flux density 4.4\,Jy at 1500\,MHz from VLA calibrator list and 6.96\,Jy at 610\,MHz from GMRT calibrator list. The flux density we obtained for PSR B1508+55 is 24.8\,mJy and the root mean square (RMS) of the image is $\mathrm{44.9\,\mu Jy}$.\\

Simultaneously with the interferometric data, phased array beam data was recorded for PSR\,B1508+55 with a frequency resolution of 24.4\,kHz, and a time sampling of 734\,ms. The processing and analysis of the data to create a dynamic spectrum are described by \citet{marthi}. We use the high signal-to-noise dynamic spectrum as the ground truth to verify our dynamic spectrum estimate from the interferometric data.\\

We also use two other observations of pulsar fields PSR J0437$-$4715 (Project Code : 06YGA01) and PSR B0031$-$07 (Project Code : 25$\_$038) to confirm and demonstrate the working of this technique on fainter pulsars in different observation conditions. The observation parameters for the three pulsars are given in Table~\ref{tab:pulsar_parameter1}.\\

\begin{deluxetable*}{lccccccccc}
\tabletypesize{\scriptsize}
\tablewidth{0pt} 
\tablecaption{Parameters for pulsar observations used in this work. We are using the beamformed data of only PSR B1508+55 for the dynamic spectrum comparison. This table shows the imaging parameters of PSR B1508+55, PSR J0437$-$4715, and PSR B0031$-$07 observations. Columns 2 to 10 describe date, subintegration time, total observation time, number of frequency channels, total bandwidth, central frequency, flux density, image rms, and the image SNR of the pulsar, respectively.\label{tab:pulsar_parameter1}}
\tablehead{
\colhead{Pulsar} & \colhead{Date} & \colhead{$t_\mathrm{subint}$}& \colhead{$t_\mathrm{obs}$}& \colhead{$N_\mathrm{chan}$} & \colhead{BW} & \colhead{$\nu_c$} & \colhead{$S_{\nu}$} & \colhead{RMS} & \colhead{SNR}\\
\colhead{} & \colhead{MJD} & \colhead{s}& \colhead{min} & \colhead{}& \colhead{MHz} & \colhead{MHz} & \colhead{mJy} & \colhead{mJy}
}
\startdata 
B1508+55& 57977&	32.2& 46&	8192&	200&	650& 24.8&	0.045&	533\\
J0437$-$4715& 53261&	16.8& 26&	128&	16&	333& 88.6&	2.56&	34.7\\
B0031$-$07& 56703&	2.01& 33&	256&	16&	325& 72.2&	2.96&	24.4\\
\enddata
\end{deluxetable*}

\section{Data analysis} 
\label{sec:data_analysis}
The data analysis and processing below were implemented with CASA (v6.5.0) and \texttt{python} scripts. We iteratively cleaned and flagged the data for radio frequency interference (RFI) and calibrated the data following standard recommendations for uGMRT data \citep{kale2021}. The iterative process allows us to remove fainter RFI and create a high$-$fidelity image.\\

\subsection{Constructing a dynamic spectrum} 
\label{sec:constructing} 

\begin{figure*}
\includegraphics[width=\textwidth]{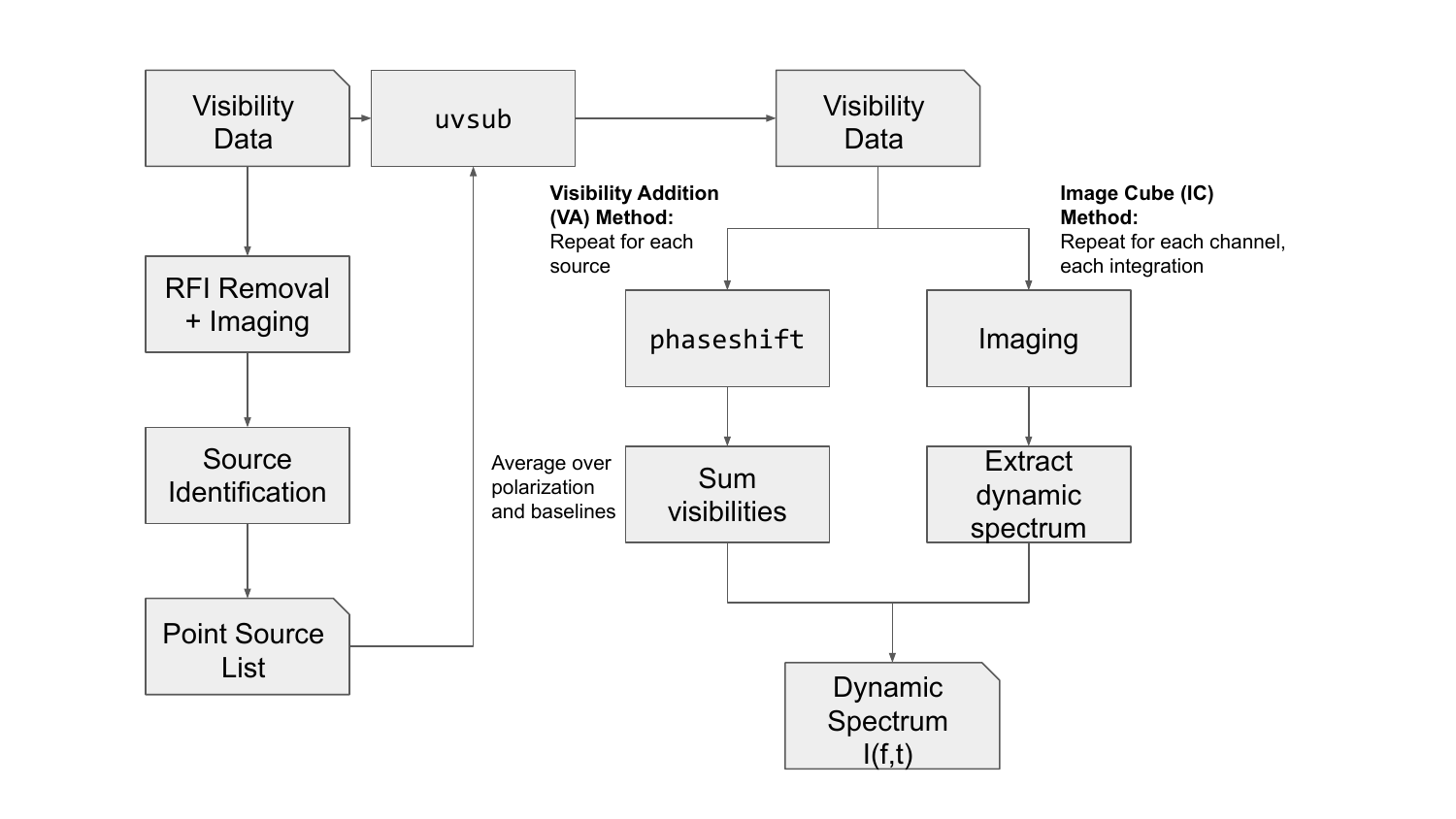}

\caption{Flowchart of constructing the dynamic spectrum from the visibility data.
\label{fig:flowchart}}
\end{figure*}

In continuum observations, the data is stored in visibilities, which are complex values as a function of time, frequency, baseline, and correlations.  As in standard processing, we create a high-fidelity image of the field with the CASA \texttt{tclean} task to invert and clean the image. Within the radio image, we identified all the point sources and we chose one source at a time to create individual dynamic spectra. We employed the CASA task \texttt{phaseshift} to align the source with the phase center.  Additionally, to remove contributions from other sources, we subtracted them using the CASA task \texttt{uvsub}. 
\texttt{uvsub} subtracts a constant flux density in frequency and time and should remove any non-scintillating flux but keep the scintillating flux in the data.  This procedure enables us to proceed with two approaches as shown in Figure \ref{fig:flowchart}.\\

\subsubsection{Image Cube (IC) Method} An image cube represents a collection of images at each frequency sample, making it a three-dimensional structure with two spatial dimensions and one frequency dimension. In this method, we utilized the phase-shifted visibility data to generate an image cube for each time sample. To obtain intensity as a function of frequency from each image cube, we employed the CASA task \texttt{specflux}. By performing this process for each time sample, we obtained the intensity as a function of both time and frequency, denoted as $I(t,f)$. While it is possible to image the entire primary beam (approx 3000$\times$3000 pixels), the volume of the data products (approximately $6\times250\,\mathrm{GB}$ for our PSR B1508+55 dataset) and the large processing time ($\sim$ month on our workstation) were prohibitive. We opted to process a smaller image (400$\times$400 pixels) which took approximately 160 hours to complete.\\

\subsubsection{Visibility Addition (VA) Method}
The image of the sky is obtained by taking a Fourier transform of the visibility in the $uv$ plane and averaging over time, frequency, and polarization. At the phase center, the phase of the source visibility is zero, which means that only the real part of the visibility contributes to the intensity. By repeatedly phase shifting the visibilities to be centered on the location of each point source detected in the image, we compute the real part of the visibilities for that source. We then construct the dynamic spectrum for that source by averaging the real part of these visibilities over polarizations and baselines. For the specific case of B1508+55, it took approximately 5 minutes to create the dynamic spectrum from the phase-shifted visibility data for a single source.\\

The visibility addition method is considerably faster than using image cubes at each snapshot and we only utilize this method for analysis going forward. In the next section, we demonstrate that the extracted dynamic spectra are equivalent.\\

\begin{figure*}
\includegraphics[width=\textwidth]{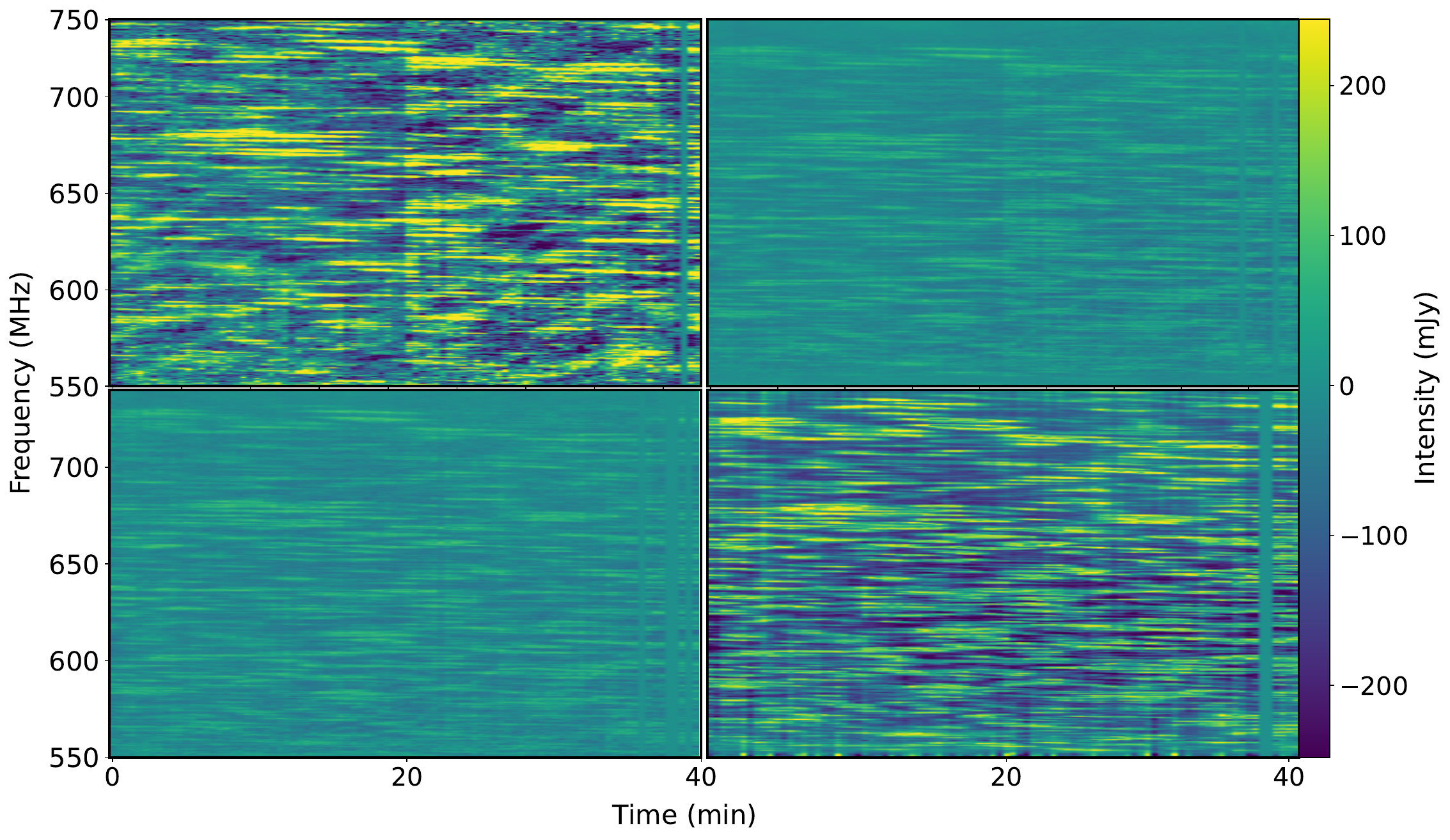}

\caption{Dynamic spectra of pulsar B1508+55 generated from different methods described in the text. The top panel presents the dynamic spectra obtained using two phase calibrator scans. The top left panel displays the dynamic spectrum derived from the image cube method, while the top right panel shows the dynamic spectrum derived from the visibility addition method. The bottom left panel shows the corrected visibility dynamic spectrum using only one phase calibrator and the bottom right panel presents the dynamic spectrum obtained through simultaneous phased array observations (as presented in Marthi et al 2020). We consider the latter as the ground truth for dynamic spectrum measurements. The dynamic spectrum is gated at the pulsar spin period to only include the `on-pulse' emission.  We note the scintles in the dynamic spectra from both the IC and VA methods are equivalent to the ground truth, albeit with lower contrast.
\label{fig:dynamic_spectra}}
\end{figure*}

\subsubsection{Verification with the Phased Array observations}
Figure \ref{fig:dynamic_spectra} shows the dynamic spectra extracted for PSR\,B1508+55 from the same data via different methods on the same gray scale. The top two panels present both the image cube (left panel) and the visibility addition (right panel) dynamic spectra of pulsar B1508+55. It is observed that the flux in the two halves of time differs in both these analyses. Upon investigation, it was realized that this discrepancy arose from utilizing two scans of the phase calibrator for calibration, with one scan conducted before the pulsar scan and the other performed after it. The later phase calibrator scan had a slightly lower amplitude, leading to higher flux in the latter half of the pulsar's dynamic spectrum.\\

To rectify this issue, we repeated the dynamic spectrum measurement using only the first phase calibrator scan. The resulting dynamic spectrum, obtained from this adjusted calibration approach, is displayed in the bottom left panel of Figure \ref{fig:dynamic_spectra}. The bottom right panel of Figure 2 shows the dynamic spectrum extracted from the simultaneous phased array beam observations \citep[as described in ][]{marthi2021}. Since this dynamic spectrum is constructed by gating the data in the on-pulse region of the pulsar, it has a much higher contrast and signal-to-noise ratio than the one constructed from visibility data.\\

Comparing the image cube dynamic spectrum and the visibility dynamic spectrum, we observe that both exhibit a similar pattern, although with differences in noise characteristics. Since constructing the dynamic spectrum from the visibility addition requires less time, we have decided to proceed with this approach.\\

Furthermore, upon comparing the visibility dynamic spectrum with the phased array dynamic spectra, we find a similar features present in both. This similarity confirms the accuracy of our method for constructing the dynamic spectrum. Further in this section, we develop the measurement methodology. The results are summarized in Section~\ref{sec:results} and the first panel of Table~\ref{tab:pulsar_table}.\\

\subsubsection{Detrending and Filtering} \label{subsec:cleaning}

\begin{figure*}
\includegraphics[width=\textwidth]{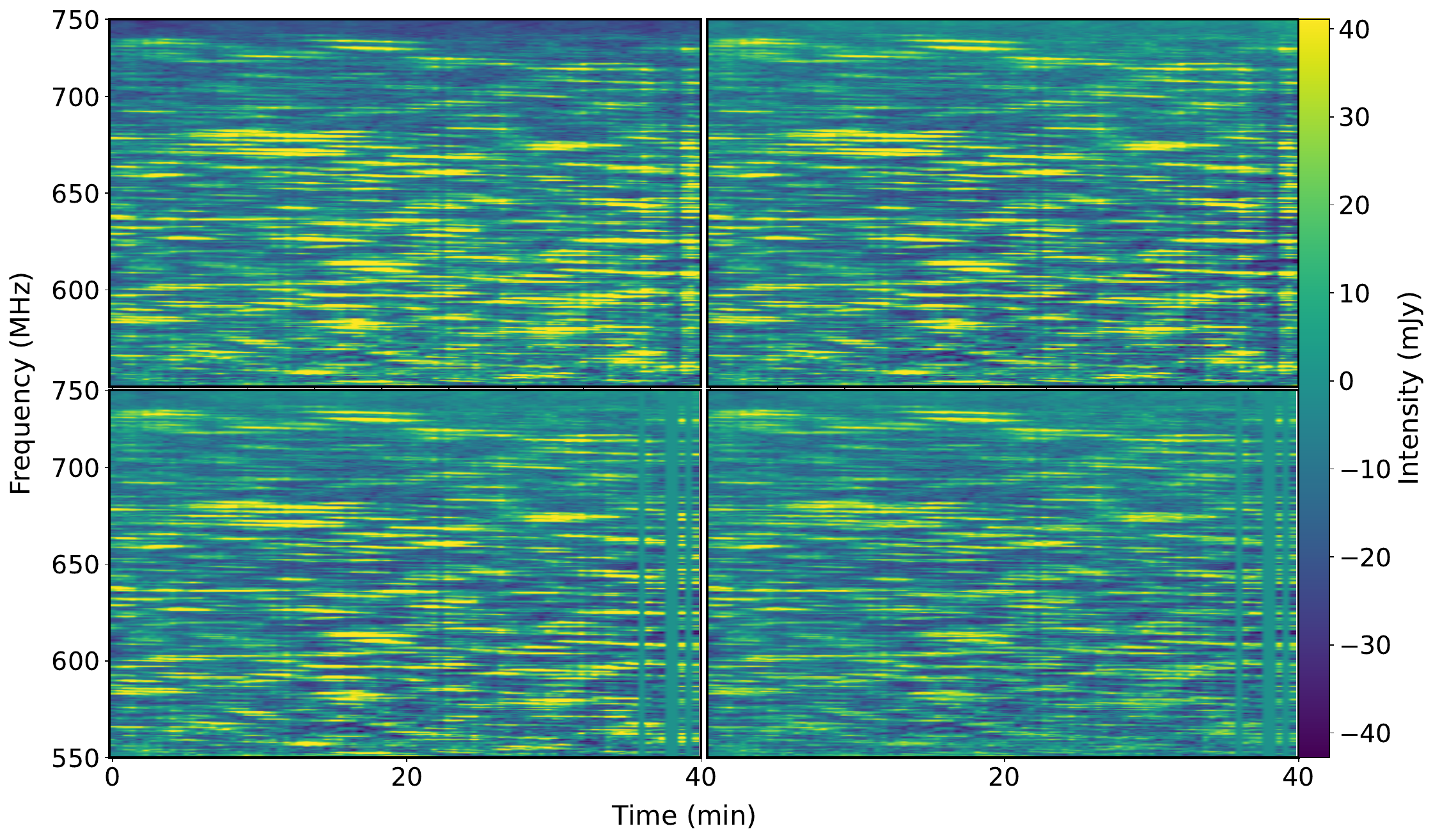}

\caption{The dynamic spectrum of PSR B1508+55 after a series of corrections, resulting in an improved representation. In the top left panel, the raw dynamic spectrum is displayed, generated directly from the visibility file. After applying a 2D detrending technique, the top right panel shows the dynamic spectrum with the large scale gradients removed. The subsequent correction involved the removal of undesirable signals, resulting in a cleaner dynamic spectrum, as shown in the bottom left panel. Finally, outlier rejection was performed to further refine the dynamic spectrum. The bottom right panel presents the dynamic spectrum after the removal of outliers, highlighting a more accurate representation of the pulsar's signal.
}
\label{fig:improved_ds}
\end{figure*}

The dynamic spectrum generated directly from the visibility file often contains artifacts and noise, making it appear ``dirty''. To enhance its quality, we have employed multiple correction techniques to reduce artefacts. In Figure \ref{fig:improved_ds}, we present the dynamic spectrum at various stages of correction. By showcasing the dynamic spectrum after each correction step, we gain insight into how and where the data was initially corrupted. This visualization allows us to track the progress of the correction process and observe the improvements achieved at each stage. The second panel of Table~\ref{tab:pulsar_table} shows the improvement in signal to noise ratio with different preprocessing steps.\\

\textbf{Detrending :} Sometimes, the dynamic spectrum may exhibit a noticeable intensity gradient along the frequency and/or time axis. This may be due to a steep spectral index for the source, residuals from the bandpass calibration, or due to a variation in gain as a function of time. This intensity gradient is evident when examining the dynamic spectrum, as shown in the top left panel of Figure \ref{fig:improved_ds}. In this particular case, the dynamic spectrum appears elevated in the middle towards positive intensity and tilted towards negative intensity along the frequency axis.\\

We remove the large-scale intensity gradients in the dynamic spectra by fitting and subtracting a 2-dimensional, 6th-order polynomial. The top right panel of Figure \ref{fig:improved_ds} shows the dynamic spectrum after subtracting the large-scale gradients.\\

\textbf{Impulsive and narrowband interference filtering :} To identify and remove any narrowband or broadband, impulsive RFI, we employ the median absolute deviation (MAD) statistics. To implement this, we calculate the median flux value for each time or frequency sample, resulting in a flux profile as a function of time or frequency. Flux values that deviate by more than 4$\times$ the median absolute deviation from the median value are considered persistent undesirable signals. The cutoff threshold of 4$\times$MAD was determined by minimizing the error in the measured scintillation parameters (described below), but our results do not depend critically on the exact value. These pixels are subsequently replaced with the median values. In the case of PSR B1508+55, about 6\% of the data were masked and replaced with the median value. The bottom left panel of Figure \ref{fig:improved_ds} showcases the dynamic spectrum after the removal of persistent undesirable signals. By comparing it to the top right panel, one can observe that it removes all the persistent low flux present at that specific time along the frequency axis.\\

\textbf{Outlier removal :} An outlier refers to an exceptionally high or low flux value that deviates significantly from the majority of other flux values at a specific time and frequency. In the analysis, flux values that exceed or fall below 5 times the MAD from the median value are identified as outliers and subsequently removed from the data. In the case of PSR B1508+55, about 2\% of the data were masked and replaced. In Figure \ref{fig:improved_ds}, the color bar scale remains consistent across all panels.\\

\subsubsection{Choice of baselines} 

When constructing the dynamic spectra from the visibilities, we have the flexibility to choose the baselines that go into it. Here we further describe two sets of choices that we use to construct and analyse the dynamic spectra. \\

\textbf{All Baselines (Auto-correlation method; AC):} 

We can achieve low noise in the dynamic spectrum by using all the baselines to construct a single dynamic spectrum and using an auto-correlation method to analyze it. However, we only get one realization of noise and RFI, making it more challenging to measure scintillation parameters and to separate scintillation from RFI and instrumental effects. We use this method to compare the dynamic spectra from the image cube and visibility addition methods to the `ground-truth' dynamic spectrum from the phased array mode (first panel of Table~\ref{tab:pulsar_table}) since the phased array beam data implicitly uses all baselines at the correlation stage.\\

\textbf{Split Baselines (Cross-correlation method; CC):}

We can also create two dynamic spectra from separate sets of baselines, selecting antennae and baselines that are less susceptible to RFI. With two separate estimates of the dynamic spectra, we can use cross-correlation to extract the scintillation parameters. We divided baselines into two equal parts, excluding the central baseline and the bad antennae. This decision was made due to the susceptibility of GMRT central baselines to heavy RFI and the specific focus on identifying point sources. Since a short central baseline was not necessary for our purposes, we excluded 91 central baselines out of a total of 435 baselines.\\

For the observation of PSR B1508+55, we identified one bad antenna along with the central baseline, which was flagged and excluded from further analysis. This resulted in a total of 315 usable baselines. To create the two dynamic spectra, we averaged 157 and 158 baselines from the total of 315 usable baselines for each set. By applying the visibility method, we obtained two distinct dynamic spectra. We cross-correlate the two spectra to measure the scintillation parameters.\\

This method reduces local RFI and creates two realizations of noise, suppressing the zero-lag peak seen in the auto-correlation function. The noise in the two sets of baselines is not completely independent as there are common antennas between the two sets, however, it is observed that the level of correlation is not significant. Consequently, the observed peak in the cross-correlation function is more likely attributed to correlated noise and the actual pulsar signal, rather than RFI.\\

This approach provides a more reliable means of studying scintillation parameters and distinguishing them from spurious signals caused by RFI. In Section \ref{sec:analysis}, we discuss the analysis methods and demonstrate the improvement in signal-to-noise ratio and scintillation parameter measurements from the different steps described above.\\

\subsection{Measuring scintillation parameters}
We fit the auto- or cross-correlation with a 2D Gaussian to measure scintillation parameter values \citep{gupta1994}. We also tried a 2D Lorentzian, but the fit was not as good. It left larger residuals compared to the 2D Gaussian. In the auto-correlation data, we mask the zero-lag values to remove the self-correlation of the noise. The functional form of two-dimensional Gaussian in a simple case of unrotated scintle is given by:
\begin{equation}
    f(t,\nu)=Ae(-\frac{t^2}{2\sigma_{t}^{2}}-\frac{\nu^{2}}{2\sigma_{\nu}^{2}}).
\end{equation}

Since our dynamic spectra have units of mJy, the unit of peak intensity ($A$) is $\mathrm{\mu Jy^{2}}$. The scintillation bandwidth, $\nu_s$, conventionally defined as the half-width at half maxima, is $\nu_s = \sqrt{2\ln2}\sigma_\nu$. The scintillation timescale, is defined as the half-width at 1/$e$ of the maxima, and is defined as $\tau_s = \sqrt{2}\sigma_t$. In practice, we fit the correlation of the dynamic spectrum with a rotated 2D Gaussian. The expressions for the functional form and the estimates of $\tau_s$ and $\nu_s$ are detailed in Appendix A.\\

The correlation SNR is defined as the ratio of the correlation amplitude to the amplitude uncertainty. In this context, the correlation peak refers to the amplitude obtained from fitting the Gaussian curve to correlation data. By evaluating the correlation SNR, we can assess the strength of the correlation relative to the background noise level.\\
We define the correlation SNR as, 
\begin{equation}
\mathrm{SNR_{cor}} = \frac{A}{\sigma_A},
\end{equation}
$A$, and $\sigma_A$ are the amplitude, and error in the amplitude of the fitted 2-D Gaussian.\\

\subsection{Calculations}

\textbf{Correlated signal :}
The scintillation of pulsar signal occurs at a distance. The full GMRT array acts like a single telescope. Scintillation will be identical since even the scattered pulsar image is unresolved. Therefore, when the signal reaches the baseline of the telescope almost all of them receive the same fluctuation in the signal. This may not be true for interferometry across very long baselines ($\sim10^4\,\mathrm{km}$. The scintillation of pulsars can be seen in the dynamic spectrum in the form of scintles and assuming its fluctuation has standard deviation $\sigma$ and the average flux density of pulsar is $S$.
The correlation signal of the dynamic spectrum of the pulsar at zero lag is
\begin{equation}
    \mathrm{signal}=\mathrm{ACF}(0,0)=<I^2>=S^2+\sigma^2.
\end{equation}

The fluctuation in pulsar signal is measured by modulation and defined as $m=\sigma/S$.\\
\begin{equation}
    \mathrm{signal}=S^2+m^2S^2=S^2(1+m^2).
\end{equation}

In our analysis, we only consider the fluctuation part of the pulsar signal. To achieve this, we subtract the mean of the dynamic spectrum, effectively setting $S=0$ for the dynamic spectrum. The corresponding auto-correlation corresponding signal is 
\begin{equation}
    \mathrm{signal}=\sigma^2=S^2m^2.
\end{equation}
\\

\textbf{Correlated Noise :}
The noise in the dynamic spectrum consists of Gaussian random noise,  baseline-dependent noise due to RFI, and fluctuation in pulsar signal. Baseline-dependent noise is typically terrestrial. Assuming that the noise in the dynamic spectra is due to Gaussian random noise ($\sigma_{n}$), baseline-dependent noise ($\sigma_{b}$), and fluctuations in the pulsar signal ($\sigma$). Then the noise in the dynamic spectrum is
\begin{equation}
\sigma_{\mathrm{DS}}=\sqrt{\sigma_n^2+\sigma_b^2+\sigma^2}.
\end{equation}

The correlation of two variables at $j^{th}$ lag is defined as

\begin{equation}
    \mathrm{Cor}(X,Y)=\frac{\sum_{i}
    x_iy_{i+j}}{\sum_i}.
\end{equation}

The variance of the product of two random variables is \citep{goodman},

\begin{equation}
   \begin{split}
       \mathrm{Var}(X,Y)=& E(X)^2\mathrm{Var}(Y)+E(Y)^2\mathrm{Var}(X) \\ &
    +\mathrm{Var}(X)\mathrm{Var}(Y).
   \end{split}
\end{equation}

If the mean of a random variable is zero and both the random variables have the same rms noise $\sigma_r$.
\begin{equation}
    \mathrm{Var}(X,Y)=\sigma^4_r.
\end{equation}

The variance of correlation is the same as averaging $N$ samples with noise $\sigma^2_r$ is 

\begin{equation}
    \mathrm{Var}(\mathrm{Cor}(X,Y))=\frac{\sigma^4_r}{N}.
\end{equation}

In the dynamic spectrum, the scintles are distributed randomly, the baseline-dependent noise is also random. Therefore the noise in the correlation is given by

\begin{equation}
    \sigma_{\mathrm{corr}}=\frac{\sigma^2_{\mathrm{DS}}}{\sqrt{N}}.
\end{equation}
\\

\textbf{Correlation SNR :} 
At zero lag, other than fluctuation in the pulsar signal, the Gaussian noise, and baseline-dependent noise also contribute to the correlation signal. The peak value of auto-correlation is equal to the quadrature sum of flux density and therefore equals to the variance of the dynamic spectra.
 
\begin{equation}
    \mathrm{max}(y_{\mathrm{cor}})=\sigma_{\mathrm{DS}}^2.
\end{equation}

For PSR B1508+55 the peak value is $\mathrm{750\,\mu Jy^2}$.\\

The autocorrelation value at zero lag is dominated by the single measurement of the Gaussian random noise. To measure the amplitude of the scintillation peak, we fitted a Gaussian model without considering the zero lag values. Later, we predict the value at zero lag from the Gaussian fitted model which removes the effect of Gaussian random noise then the max value is equal to $\sigma_b^2+\sigma^2$ which for the PSR B1508+55 comes out to be  $\mathrm{341\,\mu Jy^2}$. To remove the effect of baseline-dependent RFI, we perform cross-correlation that gives the maximum value equal to $\sigma^2$ which is $\mathrm{304\,\mu Jy^2}$. The auto-correlation maxima is always higher than the cross-correlation maxima.\\

The noise in the dynamic spectrum is $\sqrt{N}$ times more than the noise in the image where $N$ is the product of the time sample and the number of channels. Therefore the noise in the radio image corresponding to a given dynamic spectrum is $\sigma_{\mathrm{DS}}/\sqrt{N}$. The theoretical $\mathrm{SNR_{Im}}$ of a source with flux density $S$ is
\begin{equation}\label{image_snr}
    \mathrm{SNR_{Im}}=\sqrt{N}\frac{S}{\sigma_{\mathrm{DS}}}.
\end{equation}

For PSR B1508+55, the flux density calculated using the mean of the dynamic spectrum is 24.8\,mJy, and the noise is 26.4\,mJy. The total number of samples in the dynamic spectrum is 684904, which gives an image noise calculated from the dynamic spectrum of $\mathrm{34.5\,\mu Jy}$. The estimated $\mathrm{SNR_{Im}}$  is $\approx780$. However, the measured flux density of the source  24.8\,mJy, and the image noise $\mathrm{46.53\,\mu Jy}$ correspond to an image SNR of $\approx533$. 
The lower measured $\mathrm{SNR_{Im}}$ is likely due to the effects of incomplete deconvolution, the residuals of brighter sources, artifacts etc.\\

The noise in the auto-correlation of dynamic spectrum is $\sigma_{DS}^2/\sqrt{N}$ and in the cross-correlation, it is slightly more because the Gaussian random noise increase by $\sqrt{2}$ times due to baseline splitting. To calculate the $\mathrm{SNR_{cor}}$, we are fitting the 2$-$D Gaussian. Using Fisher's information matrix, we calculated the noise in the amplitude of Gaussian $\sigma_A$ (Appendix B).
\begin{equation}
    \sigma_A=\frac{\sigma_{\mathrm{cor}}}{\sqrt{\pi\sigma_{\mathrm{maj}}\sigma_{\mathrm{min}}/2}}.
\end{equation}
Here $\sigma_{\mathrm{maj}}$ and $\sigma_{\mathrm{min}}$ are spread along the major and minor axis of the fitted Gaussian model.\\

The signal (amplitude) is $S^2m^2$. Therefore the theoretical $\mathrm{SNR_{cor}}$ is
\begin{equation}\label{corr_snr}
\mathrm{TSNR_{cor}}=\sqrt{N}\frac{S^2m^2}{\sigma_{DS}^2}\sqrt{\frac{\pi\sigma_{\mathrm{maj}}\sigma_{\mathrm{min}}}{2}}.
\end{equation}

For PSR B1508+55, after considering only the fluctuation part we subtract the mean of the dynamic spectrum. The value of $\mathrm{ \sigma_{DS}=26.3\,mJy}$, $m=0.58$ and $S = 24.8\,\mathrm{mJy}$ and we get a corresponding value of $\mathrm{TSNR_{cor}} \approx 1456$.

However, the measured $\mathrm{SNR_{cor}}$ is $\approx 356$. This is because the noise in cross-correlation is not only from the cross-correlation of Gaussian random noise but also the noise from the correlation of pulsar signal fluctuation itself. After the first bright peak at the center of correlation images, it leaves faint yet brighter than the background scintillating ripples in the correlation image. Therefore it significantly increases the overall noise of the cross-correlation. To correct the theoretical noise we need to define the corrected theoretical $\mathrm{SNR_{cor}}$ expression.
\begin{equation}
\mathrm{CTSNR_{cor}} = \frac{S^2m^2}{\sqrt{\sigma_{\mathrm{DS}}^4/N+\mathrm{constant}}}\sqrt{\frac{\pi\sigma_{\mathrm{maj}}\sigma_{\mathrm{min}}}{2}}
\end{equation}
Here the constant is the constant difference between the variance of the theoretical and measured noise. The other factor is that the auto-correlation SNR is always higher than the cross-correlation SNR due to different correlation noises.\\

\subsubsection{Sensitivity limit}

The $\mathrm{SNR_{Im}}$ corresponding to $\mathrm{SNR_{cor}}$ is
\begin{equation} \label{eqn:survey_sensitivity}
    \mathrm{SNR_\mathrm{Im}}=\frac{\mathrm{SNR}_{\mathrm{cor}}\sigma_{\mathrm{DS}}}{\sigma m}\sqrt{\frac{2}{\pi\sigma_{\mathrm{maj}}\sigma_{\mathrm{min}}}}. 
\end{equation}

To calculate the minimum $\mathrm{SNR_{Im}}$ required to detect scintillation, we set the minimum $\mathrm{SNR_{cor}}$ to be 5, and the modulation should be at its maximum, which is 1.  However, it's important to note that there are other contributions to dynamic spectrum noise, which implies that the standard deviation of the dynamic spectrum noise ($\sigma_{\mathrm{DS}}$) is greater than $\sigma$.\\

The $\sigma_{\mathrm{DS}}$ can be calculated as
\begin{equation}
    \sigma_{\mathrm{DS}}=\sqrt{\sigma^2_{S}+\sigma^2}.
\end{equation}

Here $\sigma_S$ is the system noise that can be calculated from the radiometer equation.
\begin{equation}
    \sigma_S=\frac{T_{s}}{G\sqrt{2N_b\Delta\nu\Delta t}}.
\end{equation}

where $T_s$ is the system temperature, G is the antenna gain, $N_b$ is the number of baselines, $\Delta\nu$ is the channel width and $\Delta t$ is the sub-integration time.  In Section \ref{subsec:surveys}, we use these expressions to calculate the flux densities and image SNRs that are required to detect scintillation and identify pulsar candidates.

\section{Results} 
\label{sec:results}

\subsection{Analysis of Dynamic Spectra}
\label{sec:analysis}

\begin{figure*}
\includegraphics[width=0.5\textwidth]{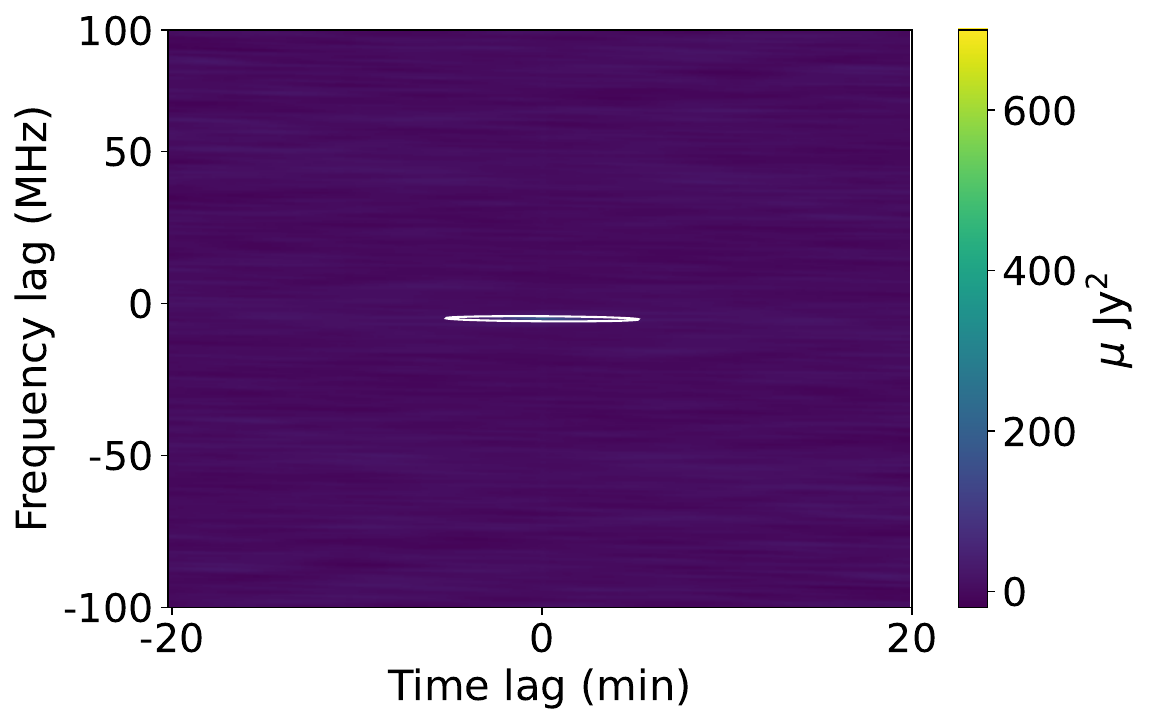}
\includegraphics[width=0.5\linewidth]{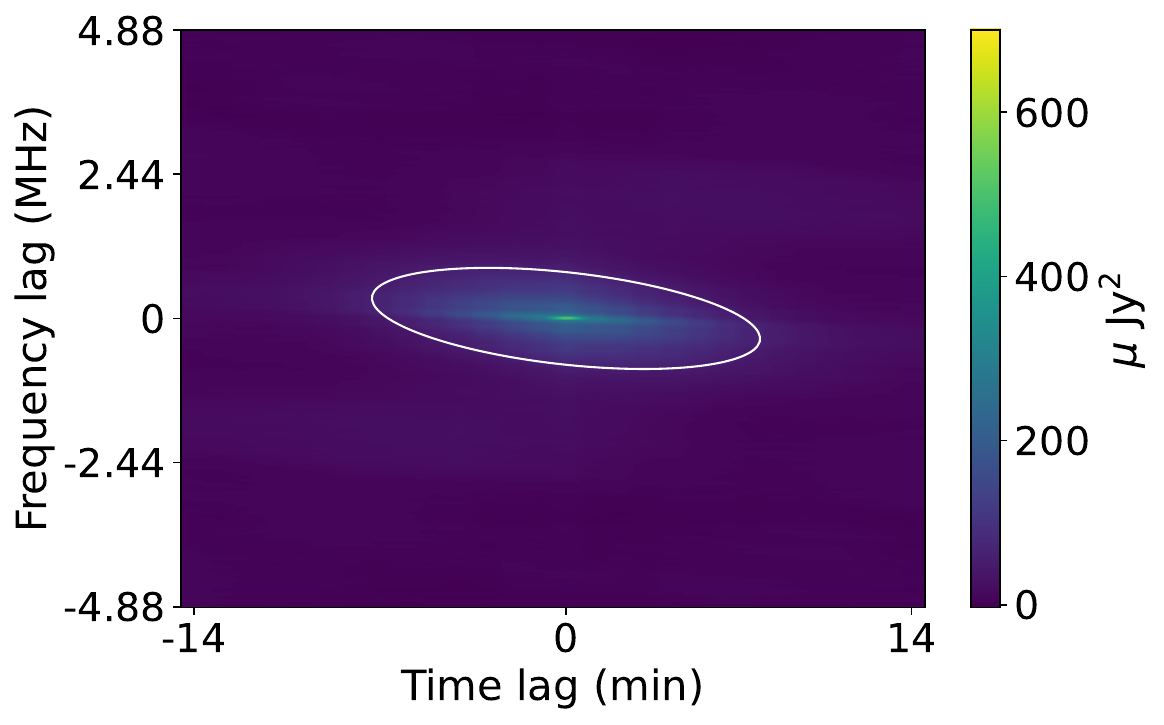}

\bigskip 
\includegraphics[width=0.5\linewidth]{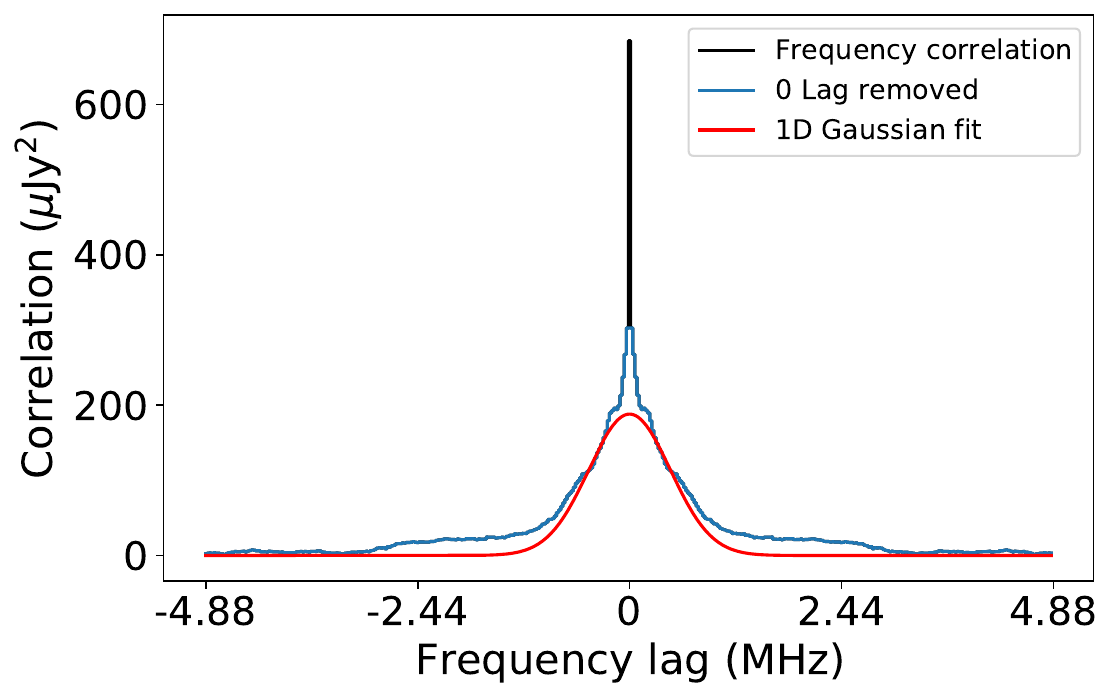}
\includegraphics[width=0.5\linewidth]{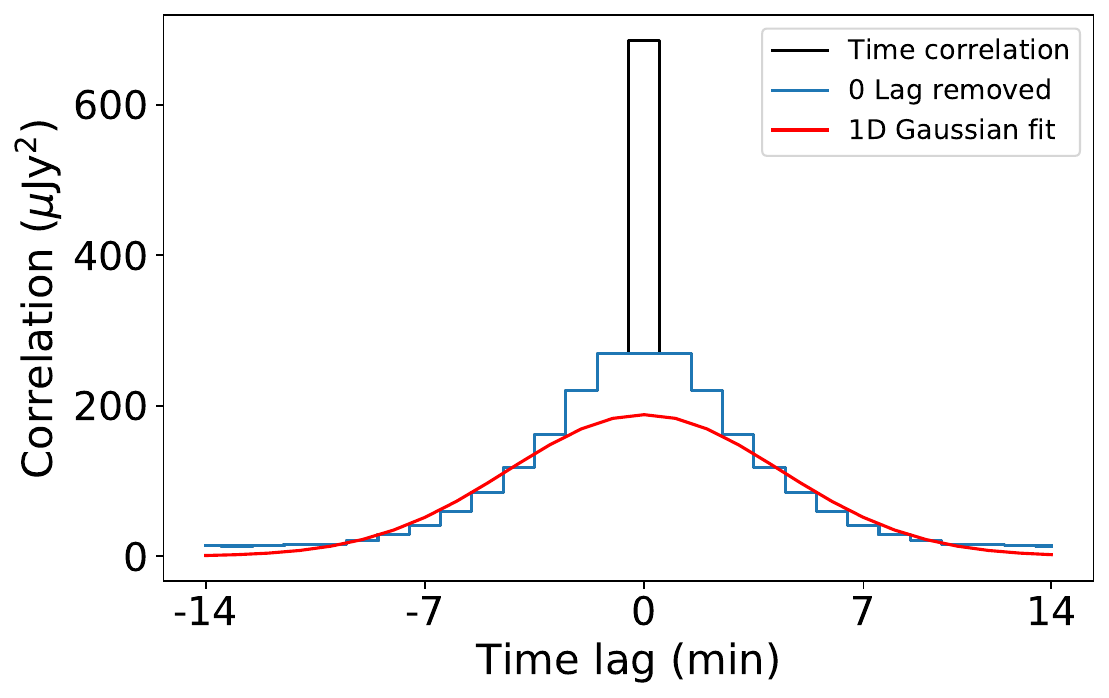}

\caption{Auto-correlation of the pulsar B1508+55 and its two-dimensional Gaussian fit. At the top panel, auto-correlation  (left panel) and zoomed-in auto-correlation (right panel) of PSR B1508+55 dynamic spectrum. In the bottom panel, the left and right panels show the auto-correlation of frequency (time) at zero time (frequency) lag with the projected one-dimensional fit, along with time and frequency correlation, respectively. The black line shows the auto-correlation values and the blue line shows the same values with the zero-lag noise peak masked. The fitted parameters are $\nu_s=594\pm1$ kHz, $\tau_s=183.6\pm0.6$ s, $\mathrm{SNR_{cor}}=386$.
\label{fig:autocorr_B1508+55}}
\end{figure*}

After constructing the dynamic spectra, we perform auto- or cross-correlation to analyze the scintillation characteristics, including the scintillation bandwidth, timescale, and amplitude. Our aim here is to detect whether scintillation exists or not. The measurement of scintillation parameters provides us with quantitative confirmation of the scintillating nature of the source. We can also use secondary spectra, but not all scintillating sources show well-defined parabolic arcs \citep{jiamusi}.\\

Scintles, representing the scintillation pattern, are randomly distributed across the dynamic spectrum. Due to the random nature of their distribution, subsequent scintle overlaps are nearly random after the initial overlap. Consequently, the auto-correlation of the dynamic spectrum peaked at small values of lags (in frequency and time) for pulsars, which provides an indication of the average size of scintles in the frequency (and time) domain.\\

\textbf{Auto-Correlation Method:}
The first panel of Table \ref{tab:pulsar_table} depicts the scintillation parameters measured from the auto-correlation of dynamic spectra from different methods. An example is shown in Figure~\ref{fig:autocorr_B1508+55}. This comparison is necessarily performed with the auto-correlation method since the phased array beam data is implicitly for the full array and we cannot use cross-correlations. The successive rows are for the image cube method, visibility addition method, visibility addition (corrected for phase calibrator flux), and the phased array beam. The visibility method (row 2 and 3) has the highest $\mathrm{SNR_{cor}}$ of scintillation. These measurements are consistent with \citet{marthi2021}.\\

\textbf{Cross-Correlation Method:}
\begin{figure*}
\includegraphics[width=0.5\textwidth]{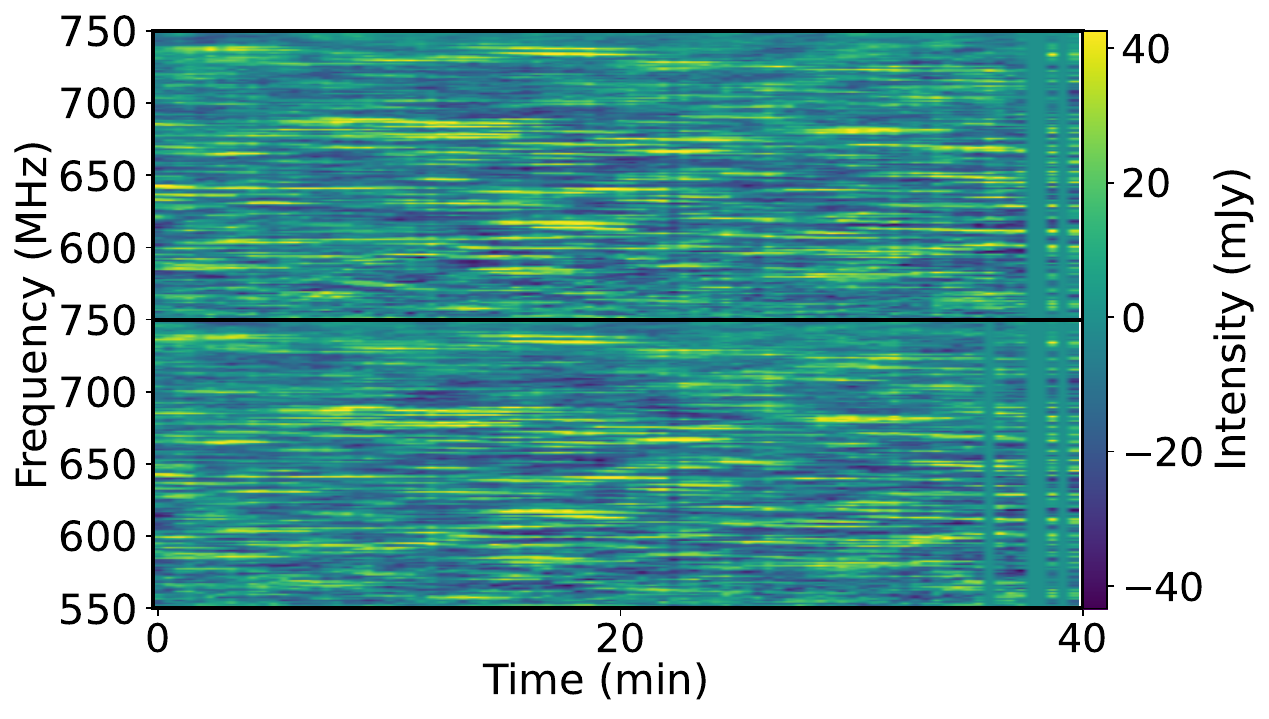}
\includegraphics[width=0.5\linewidth]{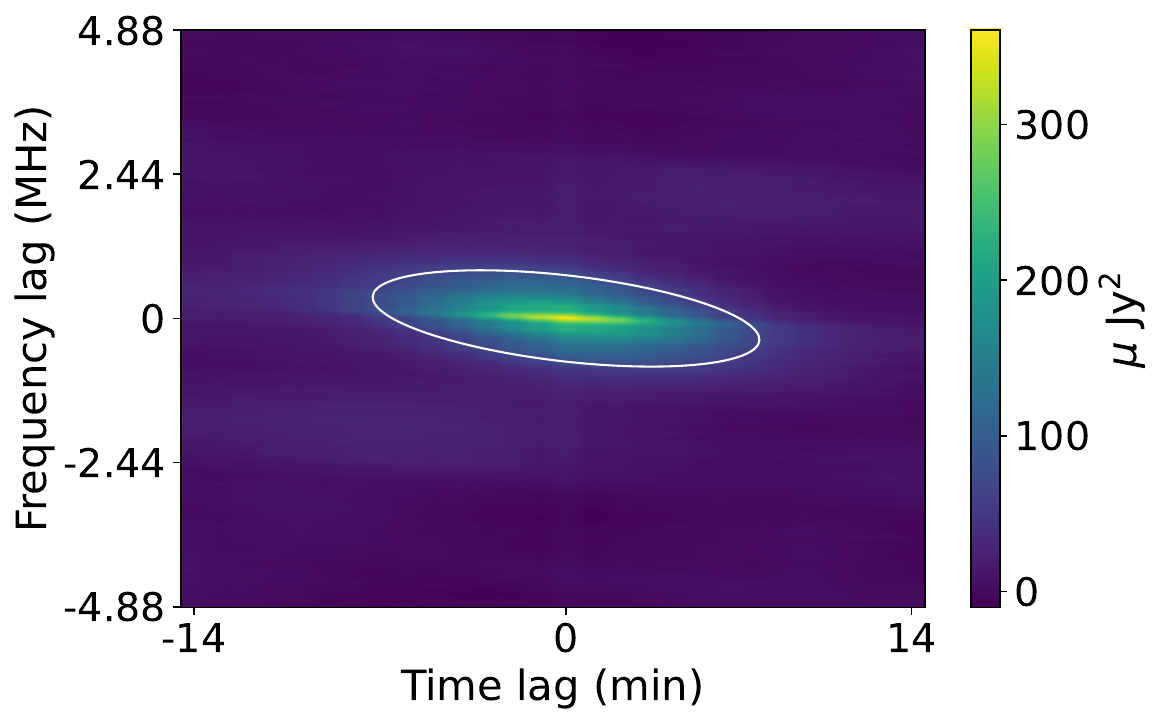}

\bigskip 
\includegraphics[width=0.5\linewidth]{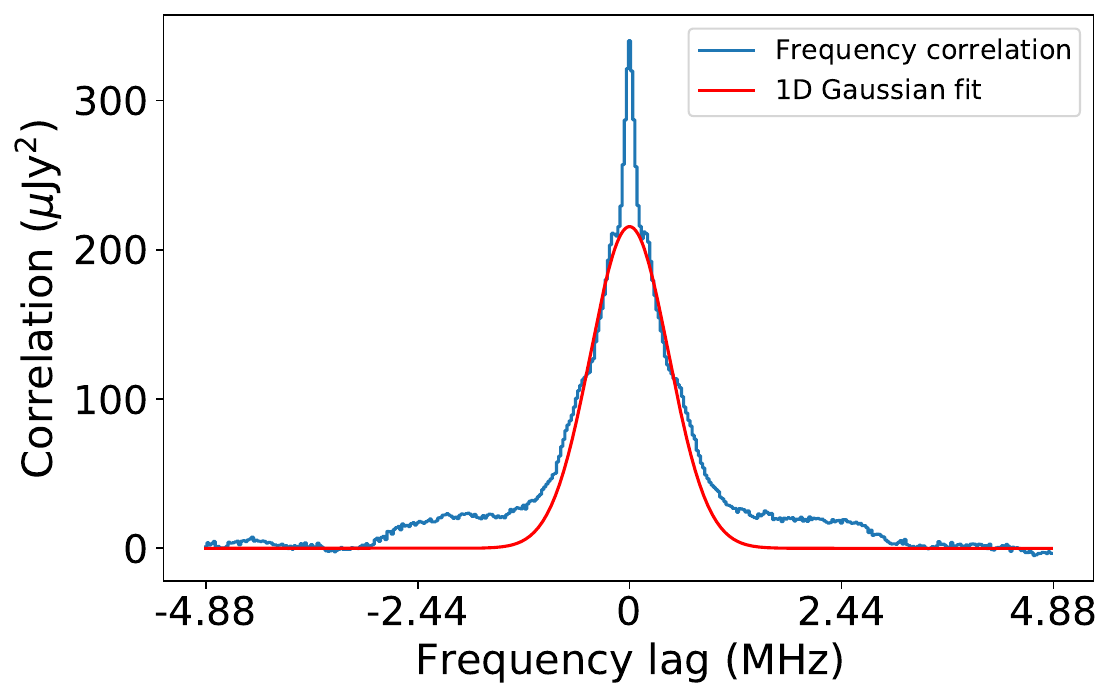}
\includegraphics[width=0.5\linewidth]{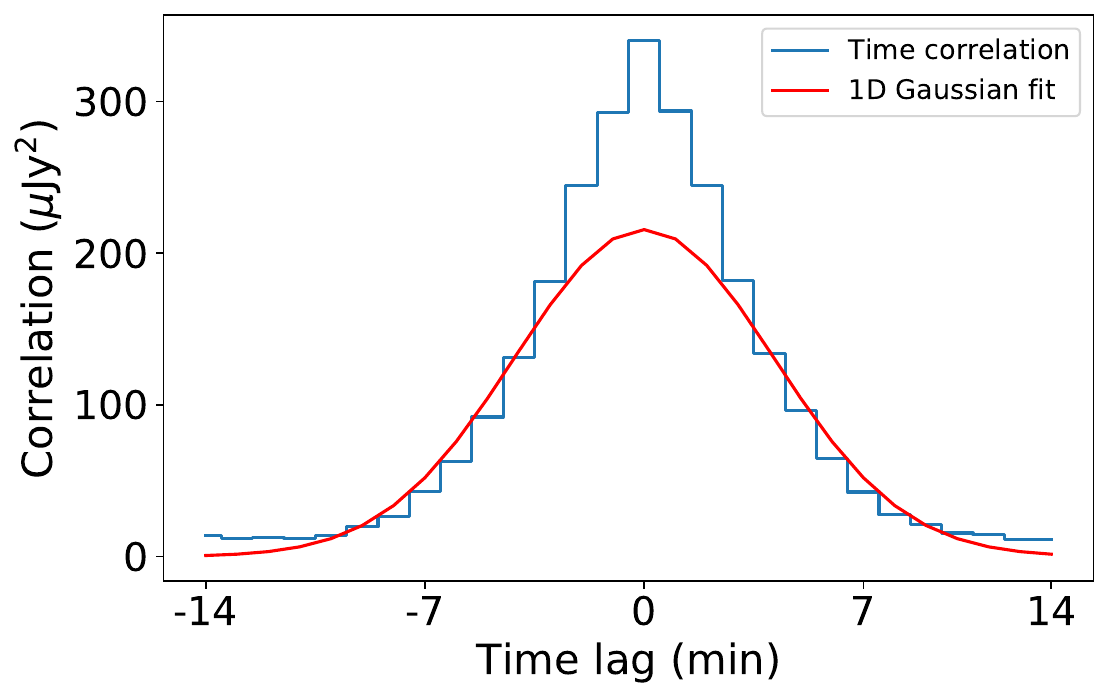}

\caption{The top left panel shows the dynamic spectra of PSR B1508+55 created from different sets of baselines (shown in sub-panels) and the top left shows the cross-correlation with the two-dimensional Gaussian fit. The bottom panels are the same as the bottom panels for Figure 4, except that the zero-lag correlation is not masked. The fitted parameters are $\nu_s=579\pm1$ kHz, $\tau_s=195.7\pm0.6$ s, $\mathrm{SNR_{cor}}$= 356.
\label{fig:cross_correlation}}
\end{figure*}

The cross-correlation of the PSR B1508+55 dynamic as depicted in Figure \ref{fig:cross_correlation} spectrum yielded the following scintillation parameters: a scintillation bandwidth of $579\pm1$\,kHz and a scintillation timescale of $195.7\pm0.6$\,s with an amplitude of $205.9\pm0.6\,\mathrm{\mu Jy^2}$. The modulation index was calculated to be 0.58, and the $\mathrm{SNR_{cor}}$ was determined to be 356.\\

The second panel of Table \ref{tab:pulsar_table} presents the scintillation parameters obtained from the cross-correlation analysis of dynamic spectra at different stages of correction. The rows correspond to the raw dynamic spectrum, the 2D detrended dynamic spectrum, the dynamic spectrum with persistent undesirable signals removed, and the dynamic spectrum with outliers removed. A noticeable enhancement in the scintillation parameters can be observed from the table. The error bars associated with the scintillation parameters decrease, resulting in more precise values. Furthermore, the $\mathrm{SNR_{cor}}$ exhibits improvement with each correction step.\\

\subsubsection{Effect of subtracting sources}

We also explored the effect of subtracting point sources from the image from the visibility data on the scintillation parameters measured. The measured scintillation parameters are presented in the third panel of Table~\ref{tab:pulsar_table}.\\

In the first test, we did not subtract any source in the visibility file and proceeded to extract a dynamic spectrum of PSR B1508+55 using the visibility addition method. In the second test, we subtracted all sources except the pulsar. In the third, we subtracted all sources, including the pulsar. Because \texttt{uvsub} subtracts a constant flux density for a source, any scintillation in the point source is maintained in the dynamic spectrum.\\

In the first method, the $\mathrm{SNR_{cor}}$ was relatively low due to the presence of contributions from other sources in the dynamic spectra. However, these contributions were relatively small, mainly within a radius of about 12\,arcsec from the central pixel, as discussed in the next section. Nonetheless, they have the potential to affect the measurement of scintillation parameters. Moreover, the second method would require a separate subtraction step for each point source in the image, adding computational effort. In the third method, a single subtraction step is sufficient for all point sources and provides equivalent results.\\

\subsubsection{Sidelobe Artefacts of Bright Scintillating Sources}

A bright source will cause scintillation artifacts in its surroundings. The sidelobes of a bright source will change with time (due to Earth rotation) and frequency (due to chromaticity). Imperfect modeling and subtraction of the dirty beam can cause large fluctuations in the dynamic spectrum, potentially resulting in inaccurate scintillation measurements. We check the extent to which the residuals from a bright source will affect its surroundings. Subsequently, we assess the extent of the scintillation effect by measuring the distance in pixels from the central pixel.\\

\begin{figure*}
 \includegraphics[width=0.53\textwidth]{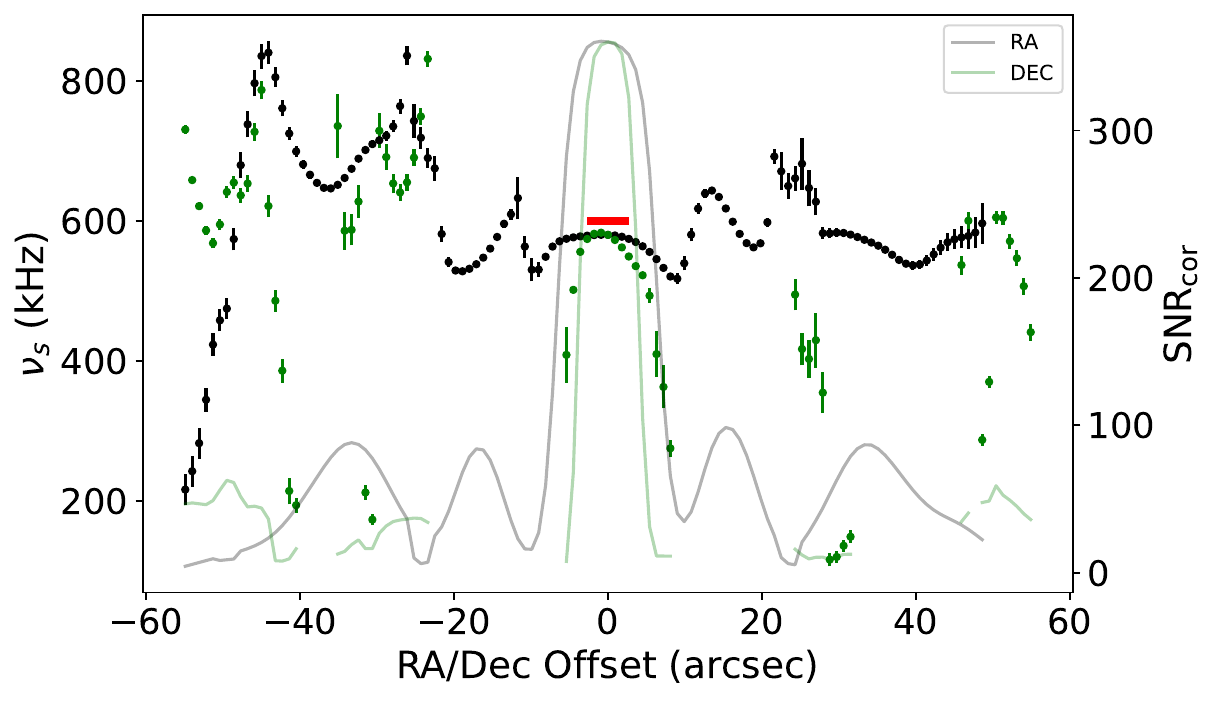}
\includegraphics[width=0.47\linewidth]{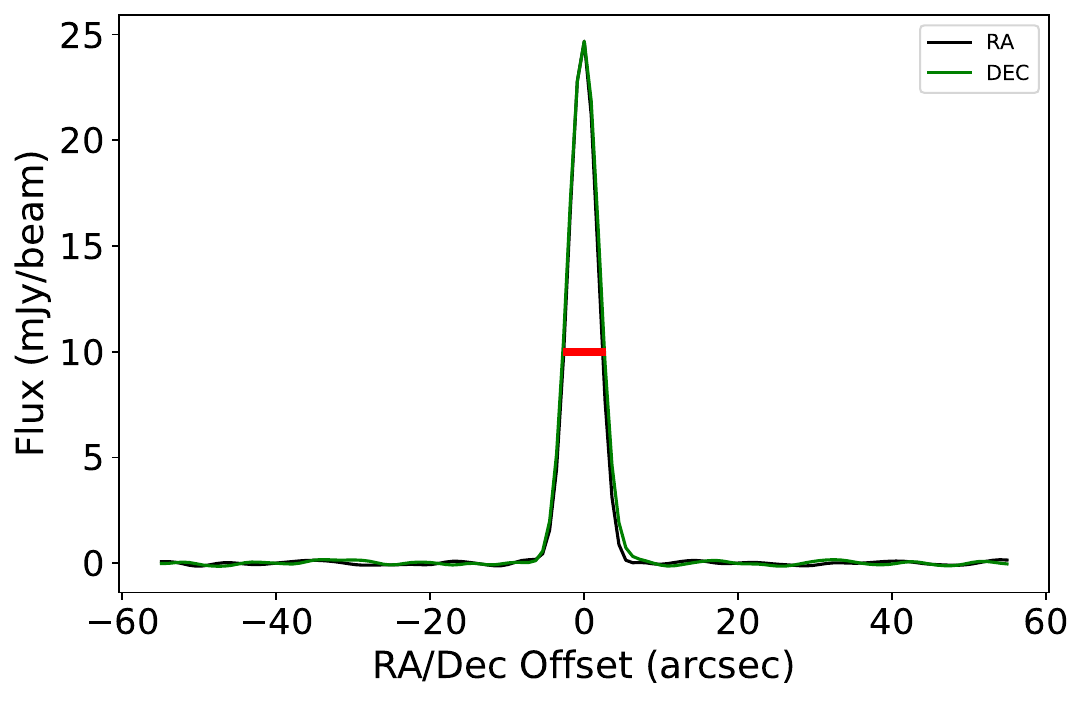}
\caption{The left panel illustrates the variation of scintillation bandwidth (dots, left axis) and $\mathrm{SNR_{cor}}$ (lines, right axis) as a function of position offset in RA and Dec around the position of PSR B1508+55. The offsets in RA, and in Dec are shown in black and green colors, respectively. The offsets are measured with respect to the centroid of PSR B1508+55; RA = 15:09:25.4 and Dec = 55:31:33.09. The vertical lines in the dots denote errors in the scintillation bandwidth. Data points with an $\mathrm{SNR_{cor}}$ less than 5 have been excluded. The right panel depicts the flux variation with respect to RA and Dec for the same .
\label{fig:scbw}}
\end{figure*}

In the case of the PSR B1508+55 observation, with a pixel size of 0.9\,arcsecs, the scintillation of the pulsar itself is measurable in the primary beam, but locations outside the primary show strong scintillation up to 1 arcmin from the central pixels as shown in Figure \ref{fig:scbw}. Outside of the primary beam, the measured value of scintillation bandwidth depends on the $uv$ distribution of the baselines and the sky location. It is important to note that the measurable range of scintillation can vary, and in general, it may extend up to a few arcminutes for different observations and sources. In implementing a scintillation based pulsar candidate search, these effects will have to be accounted for --- scintillating candidates close to bright sources will have to be analyzed carefully.\\

\subsubsection{Image Background}

\begin{figure*}
\includegraphics[width=0.48\textwidth]{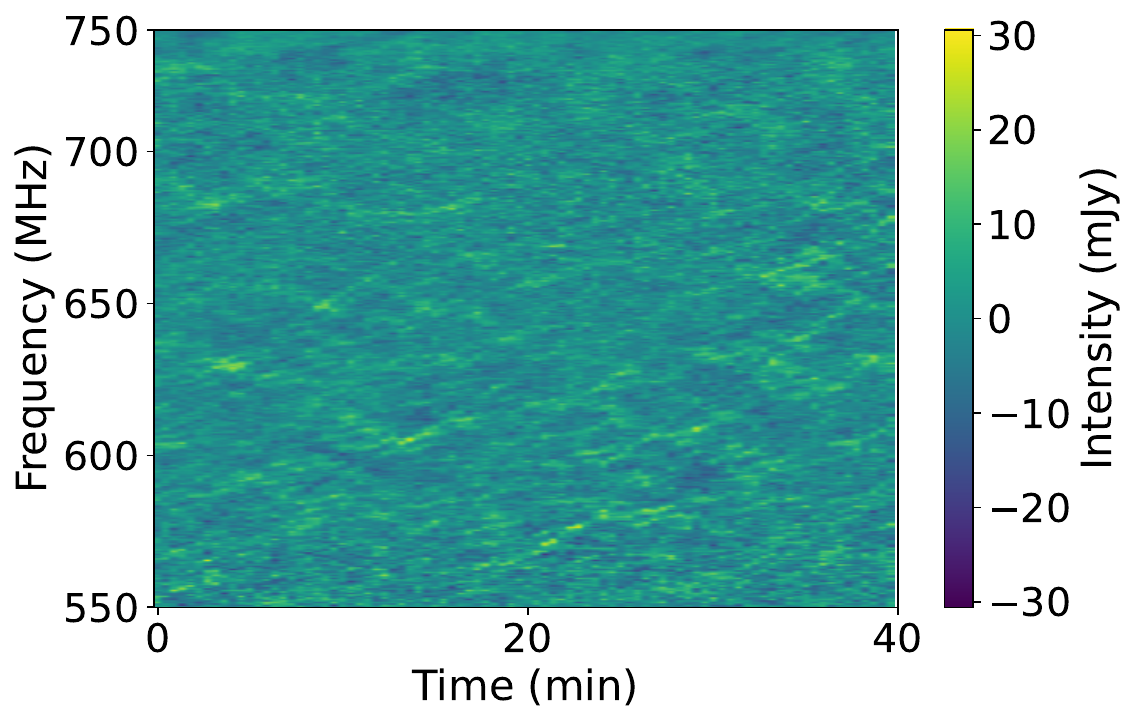}
\includegraphics[width=0.52\linewidth]{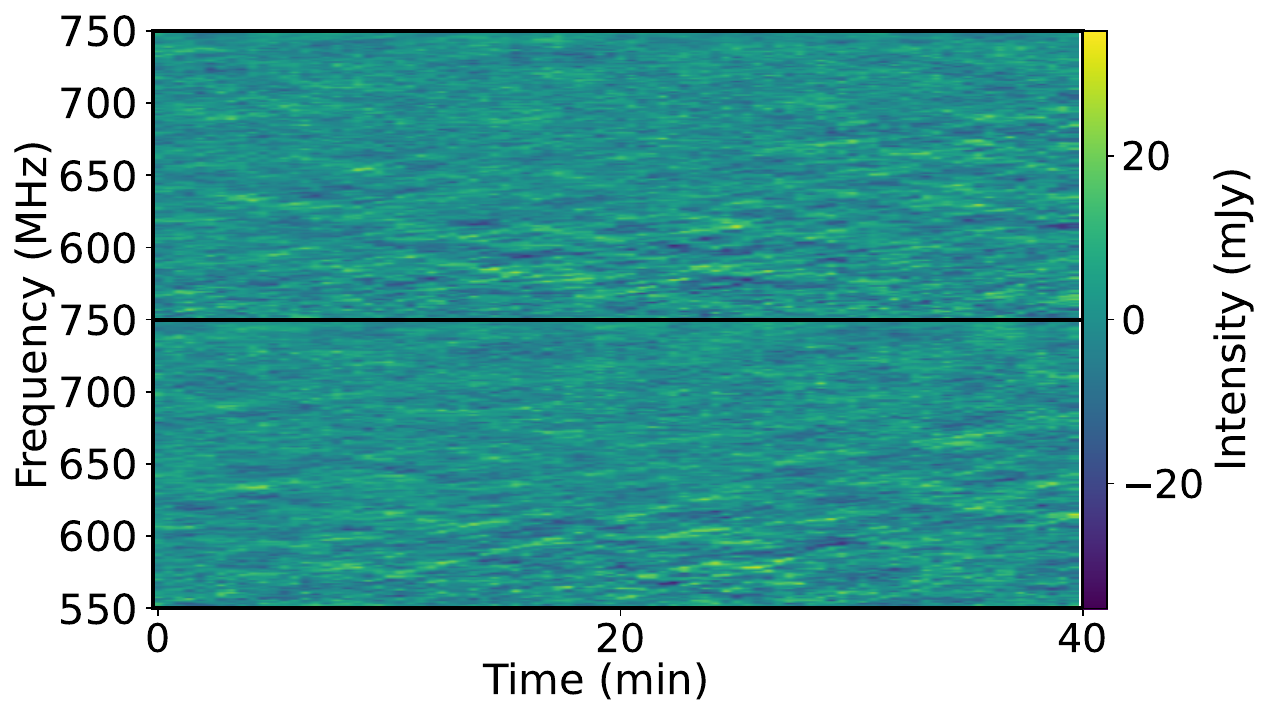}

\bigskip 
\includegraphics[width=0.52\linewidth]{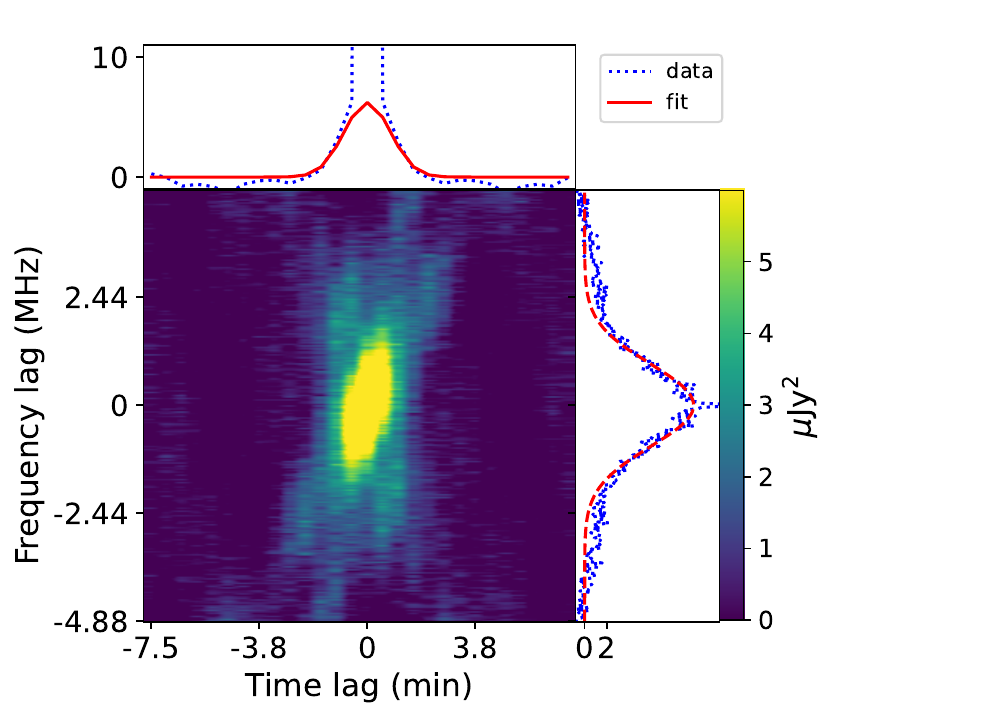}
\includegraphics[width=0.52\linewidth]{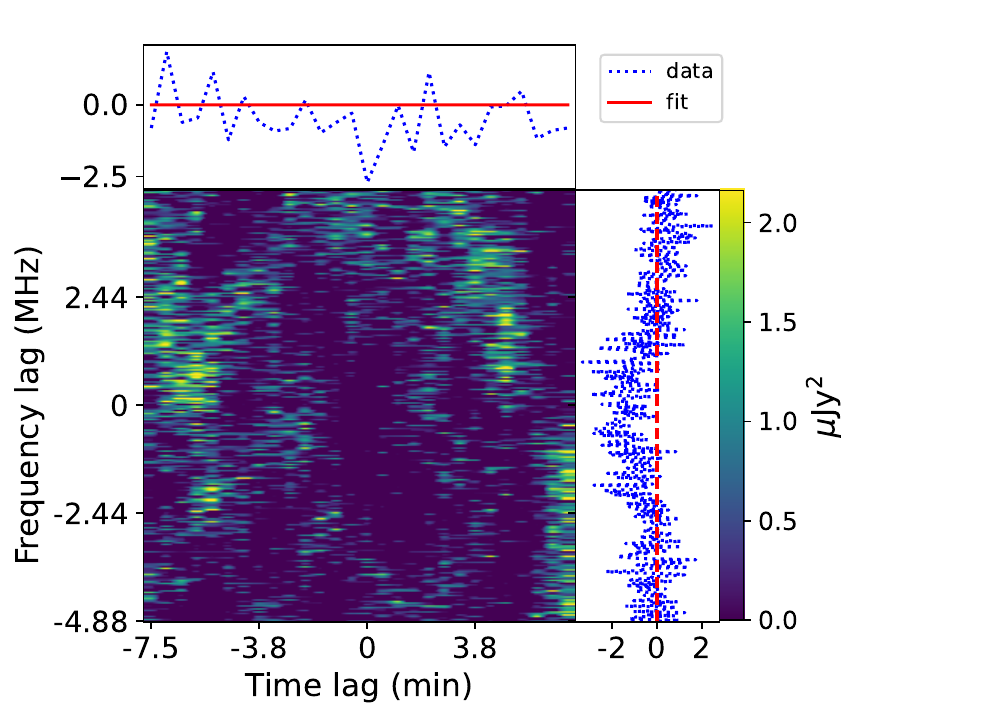}

\caption{Top panel shows the dynamic spectra of the background created from all the baselines (left) and created from different sets of baselines (right). Auto-correlation (left) and cross-correlation (right) of the background. The top sub-panel shows the frequency auto- and cross-correlation at zero time lag along with its corresponding one-dimensional Gaussian fit. The side sub-panel shows the time auto- and cross-correlation at zero frequency lag along with its corresponding one-dimensional Gaussian fit. The background is an off-source location at a separation of 108\,arcsec in RA and 216\,arcsec in Dec from PSR B1508+55.
\label{fig:image_background}}
\end{figure*}

We also examined the auto-correlation and cross-correlation of the background. In theory, the background does not scintillate, so its scintillation should ideally be zero. This analysis was conducted to verify whether any instrumental effects can affect the detection of scintillation.\\

The auto-correlation analysis unexpectedly yielded the following scintillation parameters for the background (located at $(\alpha, \delta) = 15^\mathrm{h}09^\mathrm{m}12^\mathrm{s}, +55^\circ35^\prime07^{\prime\prime}$ in J2000 coordinates): a scintillation bandwidth of $1149\pm7$\,kHz and a scintillation timescale of $54.7\pm0.4$\,s with an amplitude of $495\pm4\,\mathrm{\mu Jy^2}$, and a $\mathrm{SNR_{cor}}$ of 145. The left panels of Figure \ref{fig:image_background} show the autocorrelation of the dynamic spectrum of the background and the 2-D Gaussian fit to it.\\

The cross-correlation of the background dynamic spectra in the vicinity of PSR B1508+55 provided the following scintillation parameters: a scintillation bandwidth of $34\pm19$\,kHz and a scintillation timescale of $45\pm36$\,s with an amplitude of $59\pm40\,\mathrm{\mu Jy^2}$. The $\mathrm{SNR_{cor}}$ was measured to be 1.5. The right panels show the measured dynamic spectra, cross-correlation, and 2-D Gaussian fitted to the cross-correlation.\\

In auto-correlation analysis, both the background and the pulsar exhibit scintillation effects, making it challenging to distinguish between scintillating and non-scintillating sources. However, when we perform cross-correlation analysis, the distinction between the pulsar and the background becomes evident. The scintillation parameters of the background, characterized by large uncertainties in scintillation bandwidth and timescale, as well as a low $\mathrm{SNR_{cor}}$, strongly indicate that it is not undergoing scintillation.\\

Henceforth, we will apply the cross-correlation analysis technique to determine the scintillation parameters. The use of cross-correlation offers an additional advantage, which is elaborated upon in the correlated SNR section.\\

\subsubsection{Full Image analysis}

\centerwidetable
\begin{deluxetable*}{lcccccccc}
\tabletypesize{\scriptsize}
\tablecaption{Measured scintillation parameters and signal to noise ratios. The first of the table uses the auto-correlation (AC) method. The next three parts use the cross-correlation (CC) method. The first panel compares different methods of creating dynamic spectra. The second panel shows the measured parameters as we detrend and clean the dynamic spectra in the steps described in Subsection~\ref{subsec:cleaning}. The third panel of the table shows the effects of subtracting the sources on the measurements of scintillation parameters. The successive rows correspond to no subtraction, subtracting all but the target source, and subtracting all sources. The first three panels correspond to the PSR B1508+55 data. The last panel of the table shows the fitting parameters on the cross-correlation of the dynamic spectra of different sources from all three target fields as a verification of the technique.
\label{tab:pulsar_table}}
\tablehead{
\colhead{Sources} & \colhead{RA}& \colhead{Dec} & \colhead{$\mathrm{SNR_{Im}}$} & \colhead{$\tau_s$} & \colhead{$\nu_s$} & \colhead{$A$} & \colhead{$\theta$} & \colhead{$\mathrm{SNR_{cor}}$}\\
\colhead{} & \colhead{}&  \colhead{}& \colhead{} & \colhead{s} & \colhead{$10^2$\,kHz} & \colhead{$\mu \mathrm{Jy^2}$} & \colhead{deg} & \colhead{}
} 
\startdata
\\
\multicolumn{9}{c}{\textit{Comparison of dynamic spectra creation methods (Auto-correlation method)}}\\[0.1cm]
Image Cube& $15^\mathrm{h}09^\mathrm{m}25^\mathrm{s}$& +$55^\circ31^\prime33^{\prime\prime}$& 533& 178(1)&	7.11(1)&	173.9(9)& -6.46(6)&	354\\
Visibility& "& "& "& 207(1)&	6.61(1)&	173.9(5)&	-6.30(6)&	389\\
Visibility corrected& "& "&  "& 199(1)&	5.94(1)& 218.9(5)& -6.67(6)&	386\\
Phased array& "& "& "& 189(1)&	5.68(1)&	2770(9)&	-7.02(3)&	376\\
\hline
\\
\multicolumn{9}{c}{\textit{Comparison of pre-processing methods (Cross-correlation method)}}\\[0.1cm]
Raw data& "& "&  "& 200.3(8)&	7.31(2)&	391(1)&	-6.67(6)&	283\\
Detrended& "& "&  "& 190.7(5)&	5.67(1)&	368.8(9)&	-6.18(2)&	351\\
RFI-removed& "& "&  "& 182.2(5)&	5.49(1)&		306.1(9)&	-5.71(3)&	356\\
Outlier-rejected& "& "&  "& 195.7(6)&	5.79(1)&	205.9(6)&	-5.36(5)&	356\\
\hline
\\
\multicolumn{9}{c}{\textit{Effects of subtracting the sources (Cross-correlation method)}}\\[0.1cm]
No subtraction& "& "&  "& 189.7(5)&	5.94(1)&	199.4(5)&	-6.76(8)&	336\\
All expect PSR\,B1508+55& "& "&  "& 192.4(5)&	5.68(1)&	192.4(5)&	-6.98(4)&	341\\
All sources& "& "&  "& 195.7(6)&	5.79(1)&	205.9(6)&	-5.36(5)&	356\\
\hline
\\
\multicolumn{9}{c}{\textit{Measurements from three target fields (Cross-correlation method)}}\\[0.1cm]
B1508+55&	$15^\mathrm{h}09^\mathrm{m}25^\mathrm{s}$& +$55^\circ31^\prime33^{\prime\prime}$&  533&	195.7(6)&	5.79(1)&	205.9(6)&	-5.36(5)& 356
\\
Background&	$15^\mathrm{h}09^\mathrm{m}12^\mathrm{s}$& +$55^\circ35^\prime07^{\prime\prime}$&	0&	0.0(0)&	0.0(0)&	0.0(0)&	0.008(0)& 0
\\ 
Source1&	$15^\mathrm{h}09^\mathrm{m}22^\mathrm{s}$& +$55^\circ31^\prime26^{\prime\prime}$&  36.2&	179(2)&	5.08(3)&	12.2(1)&	-7.767(2)&	96
\\ 
Source2&	$15^\mathrm{h}10^\mathrm{m}26^\mathrm{s}$& +$55^\circ32^\prime43^{\prime\prime}$&	62.4&	54(18)&	0.3(1)&	18(5)& 	0.008(0)& 2.8
\\ 
Source3&	$15^\mathrm{h}07^\mathrm{m}57^\mathrm{s}$& +$55^\circ10^\prime33^{\prime\prime}$&	44.9&	59(24)&	0.3(1)&	14(5) &	0.008(0)&  2.44\\
 \\[0.1cm]
J0437--4715&	$4^\mathrm{h}37^\mathrm{m}16^\mathrm{s}$& --$47^\circ15^\prime02^{\prime\prime}$&  45.8&	165(4)&		6.92(2)&	37.4(9)&	-0.73(2)& 42.1
\\ 
Background&	$4^\mathrm{h}37^\mathrm{m}14^\mathrm{s}$& --$47^\circ16^\prime14^{\prime\prime}$&	0&	0.03(0)&	0.2(0)&	0(0)&	0.001(0)&  0
\\ 
Source1&	$4^\mathrm{h}39^\mathrm{m}11^\mathrm{s}$& --$46^\circ41^\prime08^{\prime\prime}$&	9.1&	108(24)&	0.5(1)&	6(2)&	0.001(0)&  3.3
\\ 
Source2&	$4^\mathrm{h}33^\mathrm{m}16^\mathrm{s}$& --$46^\circ53^\prime39^{\prime\prime}$&	20.9&	66(12)& 0.47(9)&	8(2)&	0.001(0)&  3.7
\\ 
Source3&	$4^\mathrm{h}38^\mathrm{m}47^\mathrm{s}$& --$47^\circ30^\prime56^{\prime\prime}$&	8.7&	72(30)&	0.43(7)&	5(1)&	0.001(0)& 3.9\\
 \\[0.1cm]
B0031$-$07&	$00^\mathrm{h}34^\mathrm{m}08^\mathrm{s}$& --$07^\circ21^\prime56^{\prime\prime}$ &	20.6&	0.212(1)&	18.75(2)&	50.8(3)&	-0.001(0)& 198.6
\\ 
Background&	$00^\mathrm{h}34^\mathrm{m}06^\mathrm{s}$& --$07^\circ21^\prime19^{\prime\prime}$ &	0&	4(2)&	0.9(51)&	0.05(33)&	0.0(0)& 0.16
\\ 
Source1&	$00^\mathrm{h}34^\mathrm{m}57^\mathrm{s}$& --$06^\circ57^\prime38^{\prime\prime}$&	17.2& 	4(2)& 	0.8(2)  &	2.0(6)&	0.0(0)& 3.51
\\ 
Source2&	$00^\mathrm{h}34^\mathrm{m}24^\mathrm{s}$& --$06^\circ43^\prime42^{\prime\prime}$&	8.0&	6(1)&	0.9(2)&	1.9(5)&	0.0(0)& 3.7
\\ 
Source3&	$00^\mathrm{h}34^\mathrm{m}27^\mathrm{s}$& --$06^\circ41^\prime39^{\prime\prime}$&	7.0& 9(2)&	0.9(3)&	1.4(5)&	0.0(0)& 2.8
\enddata
\end{deluxetable*}

\begin{figure*}[]
\includegraphics[width=\textwidth]{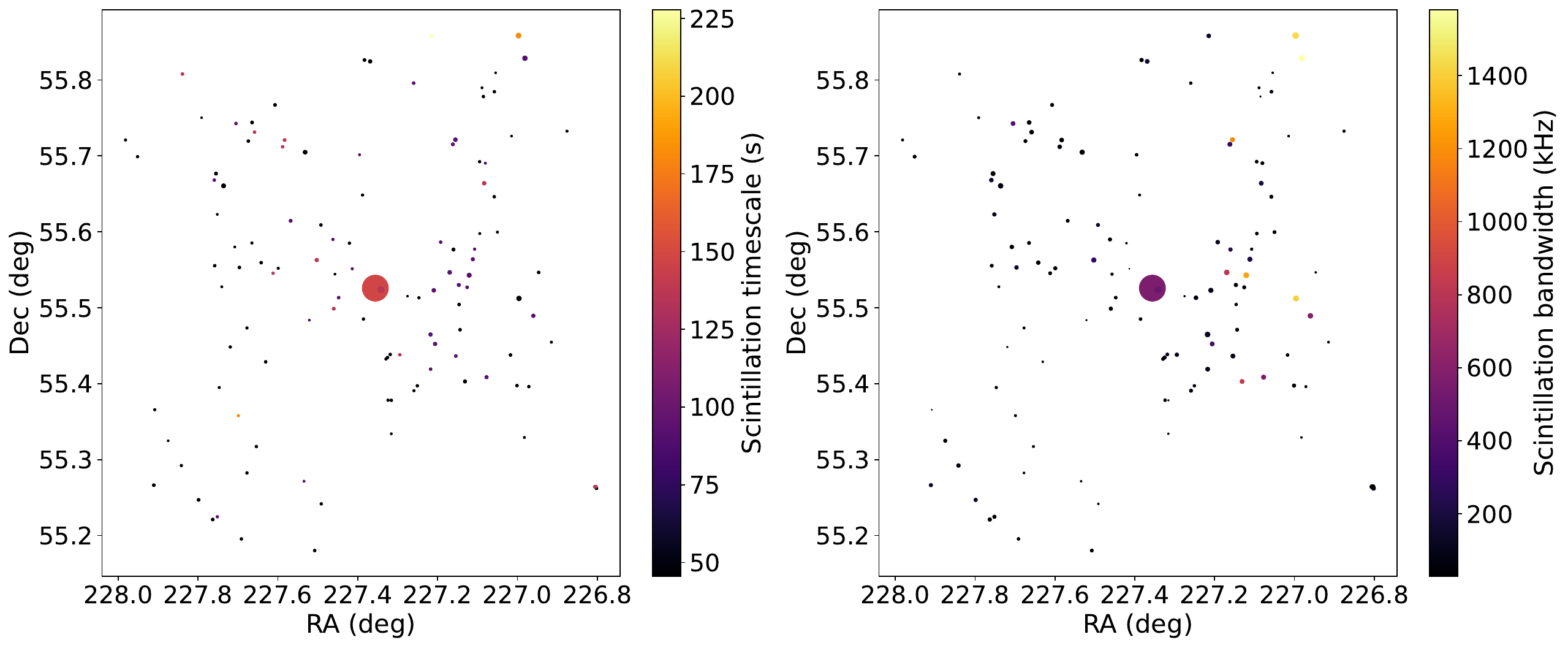}

\includegraphics[width=0.9\textwidth]{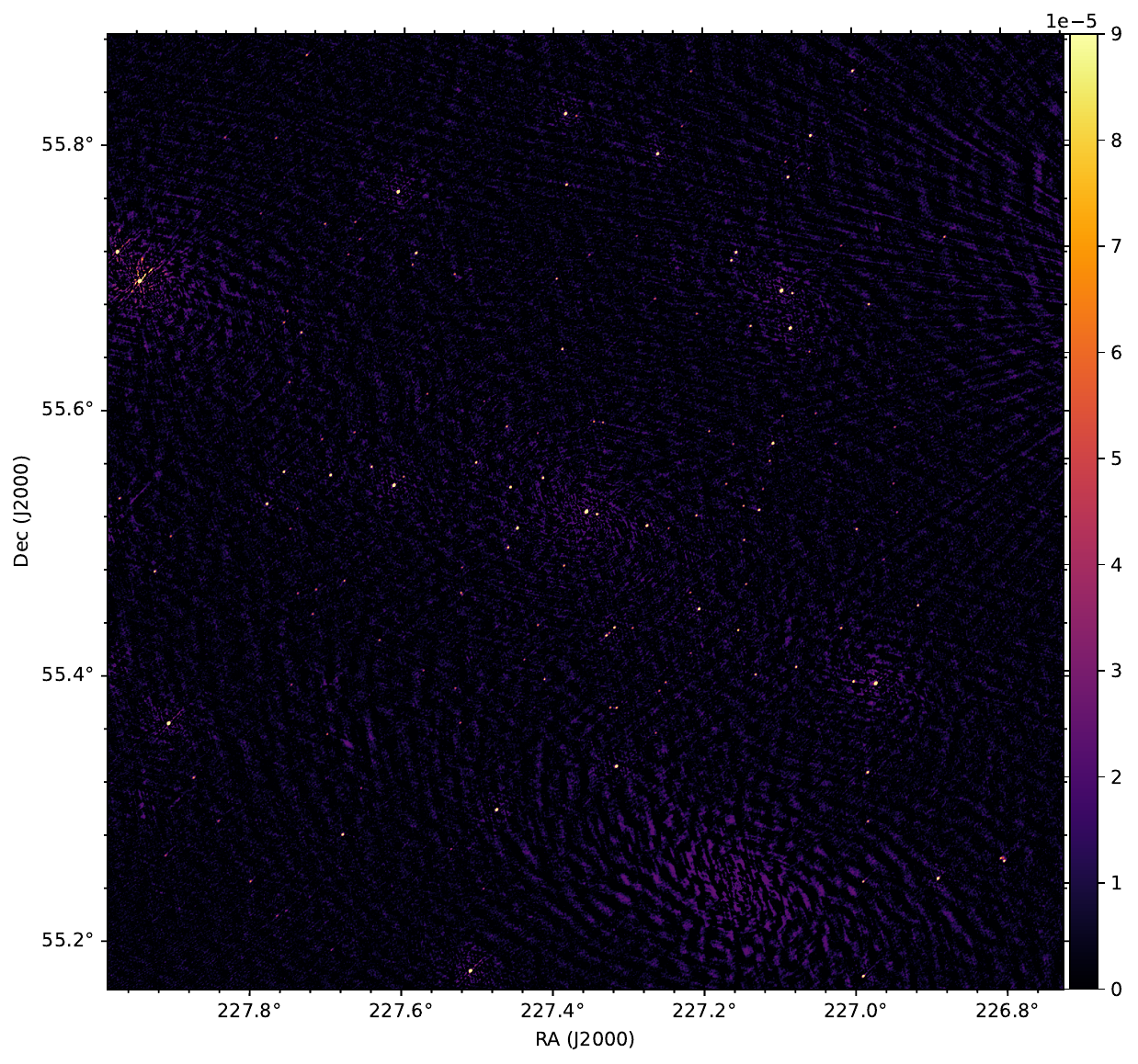}

\caption{Distribution of all the sources present in the PSR B1508+55 field as a function of sky position. In top panel, the size of the sources shows the correlation signal. The color shows the scintillation timescale (left panel) and scintillation bandwidth (right panel). The bottom panel is the radio images which shows the flux density of these sources.
\label{fig:field_analysis}}
\end{figure*}

We applied the technique to other sources present in the PSR B1508+55 field. Figure 8 shows the measured scintillation parameters for all the point sources in the image. The color bar shows the measured scintillation timescale (top left panel) and the measured scintillation bandwidth (top right panel). The size of the points sources shows the amplitude of the scintillation. The bottom panel shows the flux density of point sources and their position in the radio image. From Figure \ref{fig:field_analysis}, we clearly distinguish pulsar B1508+55 from other non-scintillating sources due to its large correlation signal.\\

We followed the identical procedure for the two additional pulsar fields and their associated sources to validate the approach. Detailed analyses of these cases are outlined in Table \ref{tab:pulsar_table}.\\

In Table \ref{tab:pulsar_table} and PSR B1508+55 field, we can observe that "source 1" shows statistically significant scintillation with very low amplitude with a scintillation timescale and bandwidth similar to that of PSR B1508+55. This source is 3 arcseconds away in RA from the PSR B1508+55. From Figure \ref{fig:scbw} we conclude that it is an artifact of the PSR B1508+55.\\

\subsection{Noise Injection Verification}

\begin{figure}
\centering
\includegraphics[width=1\linewidth]{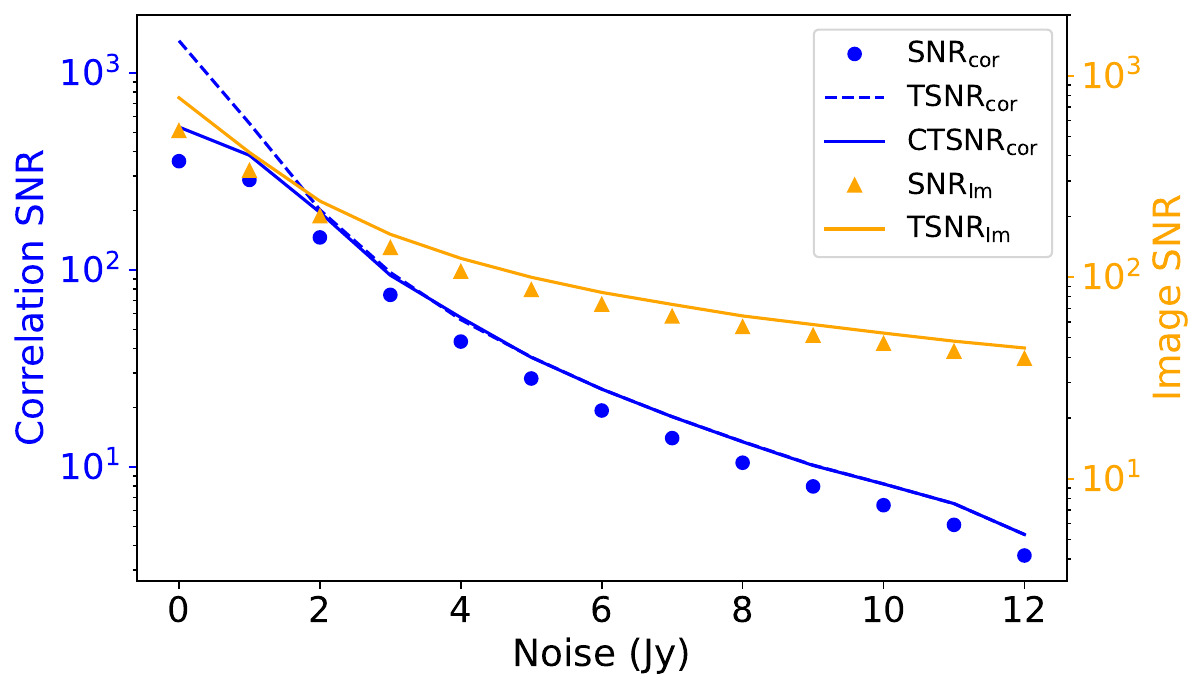}
\caption{Variation of measured $\mathrm{SNR_{cor}}$ (Blue closed circle, left axis) and $\mathrm{SNR_{Im}}$  (Orange triangle up, right axis) with the added noise and its comparison with the theoretical $\mathrm{SNR_{cor}}$. When no external noise is added, a significant difference between the theoretical and measured noise is observed. However, as we increase the magnitude of added noise, the difference between them diminishes. This trend suggests the potential to measure the $\mathrm{SNR_{cor}}$ up to its theoretical value. $\mathrm{CTSNR_{cor}}$ represents the corrected theoretical, $\mathrm{SNR_{cor}}$ which accounts for the correlation due to scintles in the dynamic spectrum and contributes to noise in the cross-correlation.
\label{fig:corrected_tsnr}}
\end{figure}

To check the sensitivity of the technique, we took PSR B1508+55 visibility data and added Gaussian random noise to the visibility data using the CASA \texttt{setnoise} tool. To each visibility, \texttt{setnoise} adds a noise 
\begin{equation}
    \sigma_{\mathrm{vis}} =\frac{\sigma}{\sqrt{n_{\mathrm{pol}}}\sqrt{n_{\mathrm{baselines}}}\sqrt{n_{\mathrm{integration}}}\sqrt{n_{\mathrm{channel}}}}.
\end{equation} 
where $\sigma_{\mathrm{vis}}$ is the additional noise in each visibility, and $\sigma$ is the total noise added in the entire dataset.
For example, if we choose $\sigma= 4\,\mathrm{Jy}$, given $n_{\mathrm{pol}}=2$, $n_{\mathrm{baseline}}=315$, $n_{\mathrm{integration}}=88$ and $n_{\mathrm{channel}}=7783$,  the noise added per visibility is $\mathrm{192.0\,\mu Jy}$. The RMS and $\mathrm{SNR_{Im}}$ of the pulsar in image after adding extra noise of 4 Jy are $\mathrm{229.6\,\mu Jy}$ and 108, respectively. The variation of $\mathrm{SNR_{cor}}$ and $\mathrm{SNR_{Im}}$ with added noise are shown in Figure \ref{fig:corrected_tsnr}.\\

\begin{deluxetable}{ccccccc}
\tabletypesize{\scriptsize}
\tablewidth{0pt} 
\tablecaption{The effect of added noise on the measurement of scintillation parameters of PSR B1508+55 in this dataset. Upto the addition of 11.0\,Jy noise, scintillation can be measured at a significance $>$ 5-$\sigma$. The value in the bracket is the 1-$\sigma$ error in the last digit.
\label{tab:noise_parameter}}
\tablehead{
\colhead{Added noise} & \colhead{$\mathrm{Noise_{DS}}$} & \colhead{$\mathrm{SNR_{Im}}$} & \colhead{$\mathrm{SNR_{cor}}$} & \colhead{$\nu_s$} & \colhead{$\tau_s$} & \colhead{$\sigma_{\mathrm{DS}}$} \\
\colhead{Jy} & \colhead{mJy} & \colhead{} & \colhead{} & \colhead{$10^2$\,kHz} & \colhead{s} & \colhead{mJy}
} 
\startdata 
0.0& 0.0 & 533& 356.3& 5.79(1) & 195.7(6)& 26.4\\
1.0& 39.8 & 338& 286.2& 5.59(1)& 186.8(7)& 48.9\\
2.0& 79.7 & 201& 146.2& 5.46(2)& 183(1)& 85.0\\
3.0& 119.5 & 140& 74.7&	5.46(5)& 182(2)& 124.0\\
4.0& 159.4 & 108& 43.3&	5.38(8)& 181(4)& 163.2\\
5.0& 199.2 & 87& 28.1&	5.3(1)&	181(7)& 203.7\\
6.0& 239.0 & 73& 19.3&	5.3(2)&	182(10) & 244.8\\
7.0& 278.9 & 64& 14.0&	5.2(3)&	185(14)& 285.4\\
8.0& 318.7 & 57& 10.5&	5.1(3)&	195(19)& 331.6\\
9.0& 358.6 & 51& 8.0&	4.8(4)&	197(25)& 375.6\\
10.0& 398.4 & 47& 6.4&	4.8(5)&	206(33)& 424.5\\
\textbf{11.0}& \textbf{438.3} & \textbf{43}& \textbf{5.1}&	\textbf{4.3(5)}&	\textbf{210(42)}& \textbf{473.7}\\
12.0& 478.1 & 39& 3.6&	3.7(5)&	198(57)& 525.9\\
\enddata
\end{deluxetable}

Our aim was to establish the theoretical sensitivity limit by determining the $\mathrm{SNR_{Im}}$ and $\mathrm{SNR_{cor}}$ for different levels of noise. For different values of added noise, we fitted a two-dimensional Gaussian function to the cross-correlation of the pulsar's dynamic spectrum in order to determine the scintillation parameters. We expected that the scintillation parameter would remain relatively unaffected by the noise since it is primarily determined by modulation, scintillation timescale, and scintillation frequency, which are not influenced by noise as shown in Table \ref{tab:noise_parameter}.\\

However, we discovered that the addition of noise had a much more significant impact on the dynamic spectra compared to the images. This resulted in faint scintillating patterns becoming obscured by the increased noise in the dynamic spectra. For the PSR B1508+55 observation, the initial noise in the dynamic spectrum was 4.1\,mJy, while the image noise was
$44.9\,\mathrm{\mu Jy}$. As we increased the noise at each step, the noise in the dynamic spectrum escalated rapidly, whereas the image noise only experienced a marginal increase. For instance, setting the noise to 0.1\,Jy resulted in the dynamic spectrum noise reaching 6.0\,mJy, while the image noise rose to $45.1\,\mathrm{\mu Jy}$.\\

The fluctuation in pulsar signal due to scintillation is 2.8\,mJy while the noise in the dynamic spectrum is 4.1\,mJy. When we cross-correlate them, due to correlated pulsar signal and uncorrelated noise the pulsar signal dominates and gives a peak and the magnitude of noise diminishes. This is especially true when the noise is much higher than the fluctuation of the signal. For instance, by setting the noise level to 1.0\,Jy using \texttt{setnoise} gives the dynamic spectrum noise 41.5\,mJy still we are able to detect the scintillation as shown in Table \ref{tab:noise_parameter}, albeit with lower $\mathrm{SNR_{cor}}$ due to longer observation and larger bandwidth. This suggests that with longer observation and larger bandwidth, we can detect the fainter scintillation signals.

The significant impact of noise on the $\mathrm{SNR_{cor}}$ compared to the $\mathrm{SNR_{Im}}$ can be explained on the basis that the $\mathrm{SNR_{cor}}$ varies with the noise as inverse square whereas $\mathrm{SNR_{Im}}$ varies as the inverse of noise.\\

 \section{Discussion} \label{sec:discussion}

\subsection{Limitations from observation data}
The detection of pulsar candidates via the technique is limited by frequency resolution, time resolution, observation time, total bandwidth, and sensitivity of the telescope. The $\mathrm{SNR_{cor}}$ decreases with the time and frequency resolution which means the finer the time and frequency resolution the higher be chances of detecting pulsar candidates.\\

\subsection{Interference and instrumental effects}

In the data analysis part of the calibration cycle, although flagging is performed to mitigate RFI, it does not completely eliminate all instances of RFI. We have shown various methods for reducing the effects of RFI and instrumental variations, however, the possibility of baseline variations manifesting as scintillation cannot be ruled out. However, we note that in such a case, scintillation parameters for many sources of the image will be similar and interdependent. We have also shown how the effects of RFI and instrumental baseline can be reduced by carefully excising short baselines that are prone to RFI.\\

 As we have seen how the presence of baseline-dependent RFI modifies the scintillation parameters in the auto-correlation and forces us to use the cross-correlation compromising $\mathrm{SNR_{cor}}$. There are other kinds of RFI present in the dynamic spectrum like correlated RFI due to ripple present in frequency and time. This kind of RFI can be observed as a repeating pattern in the correlation.\\

 The dynamic spectra from different sets of baselines gives scintillation parameters that are similar, but statistically different. The different parameters in the detrending and filtering section produce varying improvements in the different scintillation parameters. Also, the baseline splitting can not remove all the baseline dependence because the resulting two dynamic spectrum shares antennas.\\

The procedure of this technique successfully operated on the pulsar B1508+55 and its background. The results obtained from the two-dimensional Gaussian fitting of the cross-correlation of their dynamic spectrum accurately indicate that the pulsar B1508+55 is undergoing scintillation, with the scintillation bandwidth and timescale measured at 5.79(1)$\times10^2$\,kHz and 3.26(1)\,min, respectively. Furthermore, the correlation strength of the pulsar registers at 356.3, a significant value that strongly suggests the presence of pronounced scintillation.\\

\subsection{Continuum Surveys}
\label{subsec:surveys}

\begin{deluxetable}{lccccccc}
\tabletypesize{\scriptsize}
\tablewidth{0pt} 
\tablecaption{The minimum $\mathrm{SNR_{Im}}$ is required to detect scintillation in the dynamic spectra for different surveys. This assumes a modulation index of 0.4, a scintillation timescale, and a bandwidth of about 3$\times$ the sub-integration time and channel bandwidth, respectively.
\label{tab:pulsar_survey}}
\tablehead{
\colhead{Survey} & \colhead{Sensitivity}& \colhead{$t_\mathrm{subint}$}& \colhead{$t_\mathrm{obs}$} & \colhead{$N_\mathrm{chan}$} & \colhead{BW} & \colhead{$\nu_c$} & \colhead{SNR}\\
\colhead{} & \colhead{mJy/beam}& \colhead{s} & \colhead{min} & \colhead{}& \colhead{MHz}& \colhead{MHz} & \colhead{}
}
\startdata 
TGSS& 3.5&	2& 15&	256&	16.7&	147.5&	53\\
LOTSS& 0.1&	1& 480&	3934&	48&	144&	300\\
VLASS& 0.12&	0.5& 5&	125&	2000&	3000&	17\\
\enddata
\end{deluxetable}

After obtaining the calibrated radio continuum data, our technique can be implemented to identify scintillators as potential pulsar candidates within the observation field. With the growth of sensitive wide-area radio sky surveys, a search for scintillation can be easily implemented as a part of the processing or post-processing pipeline. Using Equation~\ref{eqn:survey_sensitivity} we can estimate the minimum $\mathrm{SNR_{Im}}$ or equivalently, the minimum flux density required to measure or rule out scintillation (for an assumed modulation index). Each survey will have a range of detectable scintillation bandwidths and scintillation timescales based on the survey and telescope parameters (channel width, bandwidth, integration time, observing time). We can then use our calculations to estimate the number of pulsar candidates that can be detected with a given flux density limit and modulation index. The flux density limit can be found by rearranging equation \ref{corr_snr} with a lower bound on the modulation index and an upper bound on the scintle size.
\begin{equation} 
S = \frac{\sigma_{\mathrm{DS}}}{m} \mathrm{SNR_{cor}}^{1/2} 2^{1/4} N^{-1/4} (\pi \sigma_{\mathrm{maj}}\sigma_{\mathrm{min}})^{-1/4}.
\end{equation}

Then we can calculate the image $\mathrm{SNR_{Im}}$ required to detect the pulsar candidates by substituting $S$ in equation \ref{image_snr},

\begin{equation}
    \mathrm{SNR_{Im}}=\sqrt{\frac{\mathrm{SNR_{cor}}}{m^2}\sqrt{\frac{2N}{\pi \sigma_{\mathrm{maj}}\sigma_{\mathrm{min}}}}}.
\end{equation}

Table~\ref{tab:pulsar_survey} shows the typical sensitivity numbers for existing surveys for an assumed modulation index $m = 0.4$. Subsequently, we can follow up on these pulsar candidates with sensitive, high-time-resolution timing analysis and a computationally-intensive, accelerated periodicity searches to identify possible pulsations. However, this technique can only detect scintillation if the channel width is smaller scintillation bandwidth. For imaging/spectropolarimetry focused surveys such as VLASS \citep{VLASS2020}, the channel width of 2~MHz is far larger than typical pulsar diffractive scintillation bandwidths. Low frequency surveys such as TGSS \citep{TGSSADR_2017} and LOTSS \citep{LOTSS2017} with channel widths of 65~kHz and 12.2~kHz, respectively, are better suited these scintillation searches. Future surveys can be designed with smaller channel widths to identify pulsars with even smaller scintillation bandwidths  \citep[$\lesssim10\,\mathrm{kHz}$; e.g.][]{archibald_2014}.

\subsection{Scintillation Pipeline for Survey Data}

We are also developing a comprehensive pipeline based on our technique, which automates the process of data cleaning, calibration, imaging, analysis, and classification of scintillating sources. The details of pipeline and its application surveys will be detailed in a future paper. Briefly, the pipeline ingests raw visibility data (CASA measurement set; `MS' or the uGMRT long term accumulation `LTA') and applies flags for bad baselines, bad antennas, and extreme channels. Initial flagging is performed, identifying and removing bad calibrator scans and science scans. The pipeline performs three cycles of calibration and cleaning: first, calibrating the bandpass calibrator, then the phase calibrator, and finally the science data. During the calibration process, any bad baselines are flagged. The pipeline runs two rounds of self-phase calibration on the science data, and using \texttt{PyBDSF} \citep{pybdsf}, it creates a catalog of all the point sources in the field. The pipeline then creates dynamic spectra, and correlation fits for each point source to calculate the scintillation parameters and classify the scintillating sources. A file is generated that contains metadata of the observation and various parameters associated with each source. These parameters include scan number, source number sorted according to the $\mathrm{SNR_{Im}}$, right ascension, declination, flux density, scintillation timescale and bandwidth with amplitude and associated errors, modulation index, $\mathrm{SNR_{cor}}$, dynamic spectrum noise, the goodness of correlation, and the number of pixels in the dynamic spectrum.\\

\subsection{Studying extragalactic scintillation}

We can also apply this technique to study extragalactic scintillation by forming the dynamic spectrum of extragalactic sources. So far, extragalactic scintillation has been studied by forming light curves \citep{koay}, which are sufficient to determine the strength of scintillation. However, if we want to understand more about the nature of scintillation, such as the screen distance, the scintillation bandwidth, and the scintillation timescale, we need to form the dynamic spectrum and use the standard technique we used for pulsar scintillation to analyze it. Current and upcoming surveys, like CHIME, scan the entire sky in 24 hours and operate on different frequencies. We can combine the data from various days across a range of frequencies from multiple telescopes to form the dynamic spectrum for the extragalactic compact sources that are scintillating, including interplanetary scintillation \citep{chhetri} and intra-day variability. We then down-scale the dynamic spectrum according to the scintillation regime we want to investigate. Finally, we apply our technique to determine the scintillation parameters. Extensive studies of the dynamic spectrum of different types of sources also help us constrain their scintillation parameters and identify the types of sources from the analysis of the dynamic spectrum.\\

\section{Conclusion} \label{sec:conclusion}
We have demonstrated a technique to measure scintillation parameters and identify potential pulsar candidates through scintillation from imaging radio surveys. Our technique is designed with an intended application to archival interferometric data and future sky surveys. Along with steep spectra, and compactness, our technique adds another dimension to identify pulsar candidates where computationally intensive efforts can be focussed to detect pulsations and measure timing parameters. These candidates are selected independent of the intrinsic and propagation effects that make the detection of pulsations challenging and could help detect pulsars in extremely compact, highly accelerated binaries as well as pulsars with very large duty cycles. 

We have demonstrated the technique on a very bright pulsar (PSR\,B1508+55), and two moderately bright pulsars PSRs J0437--4715 and B0031$-$07. The technique is successful in efficiently constructing the dynamic spectra for all point sources in the imaging field through stacking of phased visibilities and in measuring the scintillation parameters from the correlation of the dynamic spectrum. We have shown that the effect of RFI and instrumental ripples can be mitigated to a large degree by pre-processing of the dynamic spectrum, outlier removal, removing the short baselines, and by splitting the data into two sets of baselines. We can measure (or constrain) the scintillation parameters for all point sources in the field and identify the pulsars as the only significantly scintillating sources in their fields.

By injecting additional noise in the data, we can measure how well the technique would perform for fainter sources. We present theoretical estimates and measured values from the noise injections of how the $\mathrm{SNR_{cor}}$ and $\mathrm{SNR_{Im}}$ change with the pulsar candidate's flux density, modulation index, and scintillation parameters. With this, we can estimate the minimum flux density (or equivalently, $\mathrm{SNR_{Im}}$) for different existing and ongoing radio imaging surveys to detect scintillation in pulsar candidates. 

We are developing this technique into a modular pipeline that would automate the processing of survey data and identify scintillation candidates. The application of this pipeline to uGMRT archival data, and the estimates of identifiable pulsar candidates (based on pulsar population synthesis) will be discussed in the upcoming paper.

\bibliography{sample631}{}

\begin{thebibliography}{}
\expandafter\ifx\csname natexlab\endcsname\relax\def\natexlab#1{#1}\fi
\providecommand{\url}[1]{\href{#1}{#1}}
\providecommand{\dodoi}[1]{doi:~\href{http://doi.org/#1}{\nolinkurl{#1}}}
\providecommand{\doeprint}[1]{\href{http://ascl.net/#1}{\nolinkurl{http://ascl.net/#1}}}
\providecommand{\doarXiv}[1]{\href{https://arxiv.org/abs/#1}{\nolinkurl{https://arxiv.org/abs/#1}}}

\bibitem[{{Archibald} {et~al.}(2014){Archibald}, {Kondratiev}, {Hessels}, \& {Stinebring}}]{archibald_2014}
{Archibald}, A.~M., {Kondratiev}, V.~I., {Hessels}, J. W.~T., \& {Stinebring}, D.~R. 2014, \apjl, 790, L22, \dodoi{10.1088/2041-8205/790/2/L22}

\bibitem[{{Backer} {et~al.}(1982){Backer}, {Kulkarni}, {Heiles}, {Davis}, \& {Goss}}]{backer}
{Backer}, D.~C., {Kulkarni}, S.~R., {Heiles}, C., {Davis}, M.~M., \& {Goss}, W.~M. 1982, \nat, 300, 615, \dodoi{10.1038/300615a0}

\bibitem[{{Bhattacharyya} {et~al.}(2022){Bhattacharyya}, {Roy}, {Freire}, {Ray}, {Johnson}, {Gupta}, {Bhattacharya}, {Kaninghat}, {Ferrara}, \& {Michelson}}]{Bhattacharyya}
{Bhattacharyya}, B., {Roy}, J., {Freire}, P.~C.~C., {et~al.} 2022, \apj, 933, 159, \dodoi{10.3847/1538-4357/ac74b6}

\bibitem[{{Cameron} {et~al.}(2017){Cameron}, {Barr}, {Champion}, {Kramer}, \& {Zhu}}]{cameron}
{Cameron}, A.~D., {Barr}, E.~D., {Champion}, D.~J., {Kramer}, M., \& {Zhu}, W.~W. 2017, \mnras, 468, 1994, \dodoi{10.1093/mnras/stx589}

\bibitem[{{Camilo} {et~al.}(2015){Camilo}, {Kerr}, {Ray}, {Ransom}, {Sarkissian}, {Cromartie}, {Johnston}, {Reynolds}, {Wolff}, {Freire}, {Bhattacharyya}, {Ferrara}, {Keith}, {Michelson}, {Saz Parkinson}, \& {Wood}}]{camilo}
{Camilo}, F., {Kerr}, M., {Ray}, P.~S., {et~al.} 2015, \apj, 810, 85, \dodoi{10.1088/0004-637X/810/2/85}

\bibitem[{{Chhetri} {et~al.}(2018){Chhetri}, {Morgan}, {Ekers}, {Macquart}, {Sadler}, {Giroletti}, {Callingham}, \& {Tingay}}]{chhetri}
{Chhetri}, R., {Morgan}, J., {Ekers}, R.~D., {et~al.} 2018, \mnras, 474, 4937, \dodoi{10.1093/mnras/stx2864}

\bibitem[{Dai {et~al.}(2016)Dai, Johnston, Bell, Coles, Hobbs, Ekers, \& Lenc}]{dai}
Dai, S., Johnston, S., Bell, M., {et~al.} 2016, Monthly Notices of the Royal Astronomical Society, 462, \dodoi{10.1093/mnras/stw1871}

\bibitem[{{Faucher-Gigu{\`e}re} \& {Kaspi}(2006)}]{faucher}
{Faucher-Gigu{\`e}re}, C.-A., \& {Kaspi}, V.~M. 2006, \apj, 643, 332, \dodoi{10.1086/501516}

\bibitem[{{Frail} {et~al.}(2018){Frail}, {Ray}, {Mooley}, {Hancock}, {Burnett}, {Jagannathan}, {Ferrara}, {Intema}, {de Gasperin}, {Demorest}, {Stovall}, \& {McKinnon}}]{frail2018}
{Frail}, D.~A., {Ray}, P.~S., {Mooley}, K.~P., {et~al.} 2018, \mnras, 475, 942, \dodoi{10.1093/mnras/stx3281}

\bibitem[{Goodman(1960)}]{goodman}
Goodman, L.~A. 1960, Journal of the American Statistical Association, 55, 708, \dodoi{10.1080/01621459.1960.10483369}

\bibitem[{{Gupta}(2014)}]{yashwant}
{Gupta}, Y. 2014, in Astronomical Society of India Conference Series, Vol.~13, Astronomical Society of India Conference Series, 441--447

\bibitem[{{Gupta} {et~al.}(1994){Gupta}, {Rickett}, \& {Lyne}}]{gupta1994}
{Gupta}, Y., {Rickett}, B.~J., \& {Lyne}, A.~G. 1994, \mnras, 269, 1035, \dodoi{10.1093/mnras/269.4.1035}

\bibitem[{Hewish {et~al.}(1968)Hewish, BELL, PILKINGTON, Scott, \& COLLINS}]{hewish}
Hewish, A., BELL, S., PILKINGTON, J., Scott, P., \& COLLINS, R. 1968, Nature, 217, \dodoi{10.1038/217709a0}

\bibitem[{{Intema} {et~al.}(2017){Intema}, {Jagannathan}, {Mooley}, \& {Frail}}]{TGSSADR_2017}
{Intema}, H.~T., {Jagannathan}, P., {Mooley}, K.~P., \& {Frail}, D.~A. 2017, \aap, 598, A78, \dodoi{10.1051/0004-6361/201628536}

\bibitem[{{Kale} \& {Ishwara-Chandra}(2021)}]{kale2021}
{Kale}, R., \& {Ishwara-Chandra}, C.~H. 2021, Experimental Astronomy, 51, 95, \dodoi{10.1007/s10686-020-09677-6}

\bibitem[{Kaplan {et~al.}(2000)Kaplan, Cordes, Condon, \& Djorgovski}]{Kaplan_2000}
Kaplan, D.~L., Cordes, J.~M., Condon, J.~J., \& Djorgovski, S.~G. 2000, The Astrophysical Journal, 529, 859, \dodoi{10.1086/308307}

\bibitem[{{Kaplan} {et~al.}(2019){Kaplan}, {Dai}, {Lenc}, {Zic}, {Swiggum}, {Murphy}, {Anderson}, {Cameron}, {Dobie}, {Hobbs}, {Kaczmarek}, {Lynch}, \& {Toomey}}]{kaplan2019}
{Kaplan}, D.~L., {Dai}, S., {Lenc}, E., {et~al.} 2019, \apj, 884, 96, \dodoi{10.3847/1538-4357/ab397f}

\bibitem[{{Koay} {et~al.}(2018){Koay}, {Macquart}, {Jauncey}, {Pursimo}, {Giroletti}, {Bignall}, {Lovell}, {Rickett}, {Kedziora-Chudczer}, {Ojha}, \& {Reynolds}}]{koay}
{Koay}, J.~Y., {Macquart}, J.~P., {Jauncey}, D.~L., {et~al.} 2018, \mnras, 474, 4396, \dodoi{10.1093/mnras/stx3076}

\bibitem[{{Lacy} {et~al.}(2020){Lacy}, {Baum}, {Chandler}, {Chatterjee}, {Clarke}, {Deustua}, {English}, {Farnes}, {Gaensler}, {Gugliucci}, {Hallinan}, {Kent}, {Kimball}, {Law}, {Lazio}, {Marvil}, {Mao}, {Medlin}, {Mooley}, {Murphy}, {Myers}, {Osten}, {Richards}, {Rosolowsky}, {Rudnick}, {Schinzel}, {Sivakoff}, {Sjouwerman}, {Taylor}, {White}, {Wrobel}, {Andernach}, {Beasley}, {Berger}, {Bhatnager}, {Birkinshaw}, {Bower}, {Brandt}, {Brown}, {Burke-Spolaor}, {Butler}, {Comerford}, {Demorest}, {Fu}, {Giacintucci}, {Golap}, {G{\"u}th}, {Hales}, {Hiriart}, {Hodge}, {Horesh}, {Ivezi{\'c}}, {Jarvis}, {Kamble}, {Kassim}, {Liu}, {Loinard}, {Lyons}, {Masters}, {Mezcua}, {Moellenbrock}, {Mroczkowski}, {Nyland}, {O'Dea}, {O'Sullivan}, {Peters}, {Radford}, {Rao}, {Robnett}, {Salcido}, {Shen}, {Sobotka}, {Witz}, {Vaccari}, {van Weeren}, {Vargas}, {Williams}, \& {Yoon}}]{VLASS2020}
{Lacy}, M., {Baum}, S.~A., {Chandler}, C.~J., {et~al.} 2020, \pasp, 132, 035001, \dodoi{10.1088/1538-3873/ab63eb}

\bibitem[{{Lorimer} \& {Kramer}(2012)}]{handbook}
{Lorimer}, D.~R., \& {Kramer}, M. 2012, {Handbook of Pulsar Astronomy}

\bibitem[{Manchester {et~al.}(2005)Manchester, Hobbs, Teoh, \& Hobbs}]{manchester}
Manchester, R., Hobbs, G., Teoh, A., \& Hobbs, M. 2005, VizieR Online Data Catalog, 7245

\bibitem[{{Marthi} {et~al.}(2021{\natexlab{a}}){Marthi}, {Simard}, {Main}, {Pen}, {van Kerkwijk}, {Vanderlinde}, {Gupta}, {Roberts}, \& {Quine}}]{marthi}
{Marthi}, V.~R., {Simard}, D., {Main}, R.~A., {et~al.} 2021{\natexlab{a}}, \mnras, 506, 5160, \dodoi{10.1093/mnras/stab1970}

\bibitem[{{Marthi} {et~al.}(2021{\natexlab{b}}){Marthi}, {Simard}, {Main}, {Pen}, {van Kerkwijk}, {Vanderlinde}, {Gupta}, {Roberts}, \& {Quine}}]{marthi2021}
---. 2021{\natexlab{b}}, \mnras, 506, 5160, \dodoi{10.1093/mnras/stab1970}

\bibitem[{{Mohan} \& {Rafferty}(2015)}]{pybdsf}
{Mohan}, N., \& {Rafferty}, D. 2015, {PyBDSF: Python Blob Detection and Source Finder}, Astrophysics Source Code Library, record ascl:1502.007.
\newblock \doeprint{1502.007}

\bibitem[{{Narayan}(1992)}]{narayan}
{Narayan}, R. 1992, Philosophical Transactions of the Royal Society of London Series A, 341, 151, \dodoi{10.1098/rsta.1992.0090}

\bibitem[{Rickett(1990)}]{doi:10.1146/annurev.aa.28.090190.003021}
Rickett, B.~J. 1990, Annual Review of Astronomy and Astrophysics, 28, 561, \dodoi{10.1146/annurev.aa.28.090190.003021}

\bibitem[{{Shimwell} {et~al.}(2017){Shimwell}, {R{\"o}ttgering}, {Best}, {Williams}, {Dijkema}, {de Gasperin}, {Hardcastle}, {Heald}, {Hoang}, {Horneffer}, {Intema}, {Mahony}, {Mandal}, {Mechev}, {Morabito}, {Oonk}, {Rafferty}, {Retana-Montenegro}, {Sabater}, {Tasse}, {van Weeren}, {Br{\"u}ggen}, {Brunetti}, {Chy{\.z}y}, {Conway}, {Haverkorn}, {Jackson}, {Jarvis}, {McKean}, {Miley}, {Morganti}, {White}, {Wise}, {van Bemmel}, {Beck}, {Brienza}, {Bonafede}, {Calistro Rivera}, {Cassano}, {Clarke}, {Cseh}, {Deller}, {Drabent}, {van Driel}, {Engels}, {Falcke}, {Ferrari}, {Fr{\"o}hlich}, {Garrett}, {Harwood}, {Heesen}, {Hoeft}, {Horellou}, {Israel}, {Kapi{\'n}ska}, {Kunert-Bajraszewska}, {McKay}, {Mohan}, {Orr{\'u}}, {Pizzo}, {Prandoni}, {Schwarz}, {Shulevski}, {Sipior}, {Smith}, {Sridhar}, {Steinmetz}, {Stroe}, {Varenius}, {van der Werf}, {Zensus}, \& {Zwart}}]{LOTSS2017}
{Shimwell}, T.~W., {R{\"o}ttgering}, H.~J.~A., {Best}, P.~N., {et~al.} 2017, \aap, 598, A104, \dodoi{10.1051/0004-6361/201629313}

\bibitem[{{Wang} {et~al.}(2018){Wang}, {Han}, {Han}, {Zhang}, {Li}, {Wang}, {Han}, {Wang}, \& {Gao}}]{jiamusi}
{Wang}, P.~F., {Han}, J.~L., {Han}, L., {et~al.} 2018, \aap, 618, A186, \dodoi{10.1051/0004-6361/201833215}

\bibitem[{Wang {et~al.}(2022)Wang, Murphy, Kaplan, Klinner-Teo, Ridolfi, Bailes, Crawford, Dai, Dobie, Gaensler, Graber, Heywood, Lenc, Lorimer, McLaughlin, O'Brien, Pintaldi, Pritchard, Rea, Ridley, Ronchi, Shannon, Sivakoff, Stewart, Wang, \& Zic}]{Wang_2022}
Wang, Y., Murphy, T., Kaplan, D.~L., {et~al.} 2022, The Astrophysical Journal, 930, 38, \dodoi{10.3847/1538-4357/ac61dc}

\bibitem[{{Young} {et~al.}(2010){Young}, {Chan}, {Burman}, \& {Blair}}]{young}
{Young}, M.~D.~T., {Chan}, L.~S., {Burman}, R.~R., \& {Blair}, D.~G. 2010, \mnras, 402, 1317, \dodoi{10.1111/j.1365-2966.2009.15972.x}

\end{thebibliography}
\bibliographystyle{aasjournal}



\begin{appendix}
\section{2-d Gaussian fitting}
The 2-d rotated Gaussian centered at the origin is defined as,
\begin{equation}
\begin{split}
    & f(t,\nu)=Ae^{-(at^2+2bt\nu+c\nu^2)},\\ &
    a=\frac{\cos^2{\theta}}{2\sigma^2_{\mathrm{maj}}}+\frac{\sin^2{\theta}}{2\sigma^2_{\mathrm{min}}},\\ & b=-\frac{\sin2{\theta}}{4\sigma^2_{\mathrm{maj}}}+\frac{\sin2{\theta}}{4\sigma^2_{\mathrm{min}}},\\ & c=\frac{\sin^2{\theta}}{2\sigma^2_{\mathrm{maj}}}+\frac{\cos^2{\theta}}{2\sigma^2_{\mathrm{min}}}.
\end{split}
\end{equation}

To get the spread along the time and frequency axis. We use the equation of the rotated ellipse and then calculate its value along the $x$-axis and $y$-axis.
\begin{equation}
\sigma_t=\frac{\sigma_{\mathrm{maj}}\sigma_{\mathrm{min}}}{\sqrt{\sigma^2_{\mathrm{maj}}\sin^2{\theta}+\sigma^2_{\mathrm{min}}\cos^2{\theta}}},\,\,\,\,\,\,\sigma_{\nu}=\frac{\sigma_{\mathrm{maj}}\sigma_{\mathrm{min}}}{\sqrt{\sigma^2_{\mathrm{maj}}\cos^2{\theta}+\sigma^2_{\mathrm{min}}\sin^2{\theta}}}.
\end{equation}
\\

\section{Fisher matrix}

We use the Fisher matrix formalism to calculate the SNR of the fit with respect to the noise properties of the dynamic spectrum. The SNR does not depend on angle, therefore we are evaluating the Fisher matrices in the rotation frame of Gaussian with $x$ and $y$ are the axes along the major and minor axis of the Gaussian.\\
The elements of the Fisher matrix are calculated as
\begin{equation}
    F_{ij}=\sum_n \frac{1}{\sigma_n^2}\frac{\partial F}{\partial p_i} \frac{\partial F}{\partial p_j}.
\end{equation}

where $\sigma_n$ noise in the data is
\begin{equation}
    f(x,y)=Ae^{-x^2/2\sigma_{\mathrm{maj}}^2-y^2/2\sigma_{\mathrm{min}}^2}.
\end{equation}
The partial derivatives with respect to $A, \sigma_\mathrm{maj},$ and $\sigma_\mathrm{min}$ are,
\begin{equation}
\begin{split}
    &\partial f/\partial A= e^{-x^2/2\sigma_{\mathrm{maj}}^2-y^2/2\sigma_{\mathrm{min}}^2},\\
    &\partial f/\partial \sigma_{\mathrm{maj}}= \frac{Ax^2}{\sigma_{\mathrm{maj}}^3}e^{-x^2/2\sigma_{\mathrm{maj}}^2-y^2/2\sigma_{\mathrm{min}}^2},\\
    &\partial f/\partial \sigma_{\mathrm{min}}= \frac{Ay^2}{\sigma_{\mathrm{min}}^3}e^{-x^2/2\sigma_{\mathrm{maj}}^2-y^2/2\sigma_{\mathrm{min}}^2}.
\end{split}
\end{equation}

\begin{equation}
   \begin{split}
       F_{AA}=&\frac{\pi\sigma_{\mathrm{maj}}\sigma_{min}}{\sigma_n^2},\;\; F_{\sigma_{\mathrm{maj}}\sigma_{\mathrm{maj}}}=\frac{A^2\pi\sigma_{min}}{4\sigma_{\mathrm{maj}}\sigma_n^2},\\
&F_{\sigma_{min}\sigma_{min}}=\frac{A^2\pi\sigma_{\mathrm{maj}}}{4\sigma_{\mathrm{min}}\sigma_n^2}.
   \end{split} 
\end{equation}
\begin{equation}
   \begin{split}
       &F_{A\sigma_{\mathrm{maj}}}=F_{\sigma_{\mathrm{maj}} A}=\frac{A\pi\sigma_{\mathrm{min}}}{2\sigma_n^2},\\   &F_{A\sigma_{\mathrm{min}}}=F_{\sigma_{\mathrm{min}} A}=\frac{A\pi\sigma_{\mathrm{maj}}}{2\sigma_n^2},\\   &F_{\sigma_{\mathrm{maj}}\sigma_{\mathrm{min}}}=F_{\sigma_{\mathrm{min}}\sigma_{\mathrm{maj}}}=\frac{A^2\pi}{4\sigma_n^2}.
   \end{split} 
\end{equation}

The covariance matrix is the inverse of the Fisher matrix. and the standard deviation in the amplitude is 
\begin{equation}
    \sigma_A=\sigma_n\sqrt{\frac{2}{\pi\sigma_{\mathrm{maj}}\sigma_{\mathrm{min}}}}.
\end{equation}

\end{appendix}

\end{document}